\documentclass[12pt]{article}

\usepackage{epsf}
\usepackage{latexsym,euscript}
\usepackage[dvips]{graphicx}
\usepackage{graphbox}
\usepackage{amsmath}
\usepackage{amsfonts}
\usepackage{amssymb, epsfig}

\usepackage{color,soul}

\begin{document} 
	
\begin{center}
{\Large \textbf{Lattice gauge theories in the strong coupling and static limits as a sign-problem-free Ising model}}

\vspace*{0.6cm}
\textbf{B. All\'es\footnote{email: alles@pi.infn.it}} \\
\vspace*{0.1cm}
\centerline{\it INFN Sezione di Pisa, Largo Pontecorvo 3, 56127 Pisa, Italy}
\vspace*{0.3cm}
\textbf{O.~Borisenko\footnote{email: oleg@bitp.kiev.ua}} \\
\vspace*{0.1cm}
\centerline{\it INFN Gruppo Collegato di Cosenza, Arcavacata di Rende, 87036 Cosenza, Italy and} 
\centerline{\it N.N.Bogolyubov Institute for Theoretical Physics,} 
\centerline{\it National Academy of Sciences of Ukraine, 03143 Kiev, Ukraine}
\vspace*{0.3cm}
\textbf{A. Papa\footnote{email: papa@fis.unical.it}} \\
\vspace*{0.1cm}
\centerline{\it Dipartimento di Fisica, Universit\`a della Calabria and}
\centerline{\it INFN Gruppo Collegato di Cosenza, Arcavacata di Rende, 87036 Cosenza, Italy}
\vspace*{0.3cm}
\textbf{S.~Voloshyn\footnote{email: s.voloshyn@bitp.kiev.ua}} \\
\vspace*{0.1cm}
\centerline{\it N.N.Bogolyubov Institute for Theoretical Physics,} 
\centerline{\it National Academy of Sciences of Ukraine, 03143 Kiev, Ukraine}
\end{center}

\vspace*{1cm}

\begin{abstract}
  The effective action of the $SU(N)$ Polyakov-loop model in the strong coupling region and in the static limit for the quark determinant
  can be mapped onto the Ising model in any dimensions, with the Ising variables attached on the links of the lattice. We use this reformulation
  to study the finite temperature $SU(2)$ lattice gauge theory at finite baryon density. 
\end{abstract}

\newpage

\section{Introduction} 
\label{intro} 

\subsection{Motivation} 

Dual formulations of gauge models coupled to dynamical matter remain one of the few approaches which allow to overcome the sign problem and study QCD at finite baryon density. At present, most  dual formulations with a positive Boltzmann weight have been derived in the static approximation for the full quark determinant~\cite{Gattringer11,Gattringer12a,Philipsen12,Borisenko20,Borisenko19}. 
But even in this case the extraction of long-distance quantities remains a difficult task. For this reason many numerical simulations that in the past have employed dual formulations restrict the study to local quantities like the free energy density, the quark density and the quark condensate~\cite{Gattringer12a,Philipsen12,Borisenko19}.

Clearly it would be very desirable to be able to extend those studies by including the computation of long-distance quantities like the Polyakov-loop correlation function. This achievement would allow to extract the string tension (in the confining phase) and the screening chromo-electric and chromo-magnetic masses (in the high-temperature deconfining phase). So far, there are very limited results on the behavior of such masses (for a general review on screening masses we refer to~\cite{Bazavov_2020teh}). 
In Ref.~\cite{bonati2018} these masses have been computed in lattice QCD with imaginary baryon chemical potential. Reverting to the dual formulation, preliminary results obtained by some of us can be found in~\cite{masses_lat21}, where
the screening masses were calculated with static quark determinant and in the presence of a real baryon chemical potential. Full results regarding the model in~\cite{masses_lat21} will appear in~\cite{masses_su3}. 

In the present paper we rewrite the action of the Polyakov-loop model with static determinant after having mapped it onto an Ising model, where the Ising spins are attached to the links of the lattice. This construction is valid for any dimension and for any gauge group $SU(N)$. Also, we derive the corresponding representations for various observables including the long-awaited Polyakov-loop correlation function. Since such reformulation allows for relatively manageable simulations, we also derive in terms of the new Ising variables the expressions of the string tension and the screening masses in addition to other more standard local observables.

To test the efficiency of our representation, in this first paper we simulate the $SU(2)$ case and compute various quantities among which the above ones as well as the quark condensate, particle density, two-point correlation functions of the Polyakov loops and the corresponding screening masses. This is done both in the pure gauge model and in the theory with one flavor of staggered fermions. 

The study of various versions of the $SU(2)$ Polyakov-loop models has a long history. One of the first such models was derived in the strong coupling approximation of the pure gauge theory in Ref.~\cite{polonyi_82}. 
Leading corrections to that model have been obtained in~\cite{philipsen_11}, 
together with contributions from the adjoint characters to the effective Polyakov-loop model couplings. Simulations of this improved model show good agreement with simulations of the full pure gauge $SU(2)$ theory~\cite{philipsen_11}. 

The static approximation to the full quark determinant, either with Wilson or with staggered fermions, is widely used to study $SU(N)$ models at finite density. The major simplification emerging in this approximation is that the spatial interaction between fermions is neglected and the remaining determinant (which contains the mass term and interaction in the temporal direction) can be calculated exactly. 
The result depends on the Polyakov loops and the effective theory can be studied both analytically, using some approximate methods, or numerically after dualization 
(this gives a positive Boltzmann weight even in the presence of baryon chemical potential). The static approximation can be justified in two cases: 1) large mass and/or large chemical potential; 2) on anisotropic lattices when the hopping in the spatial direction is suppressed by the anisotropy parameter ${a_t}/{a_s}\ll1$, where $a_t,a_s$ are the temporal and spatial lattice spacings.
Leading corrections to the static determinant with Wilson fermions have been calculated in the hopping parameter expansion in~\cite{philipsen_14}. 
One expects similar corrections to the static contribution also in the case of staggered fermions.

The phase diagram of the $SU(2)$ model is well known in the pure gauge case. In the full theory with dynamical quarks there is a conjectured phase diagram, see, {\it e.g.}, Ref.~\cite{kogut_02}. Agreement with this tentative diagram has been obtained in Refs.~\cite{diquark_cond_01,tomeu_06}, even though the simulations were done on rather small lattices. The central quantity in the description of the critical behavior is played by the diquark condensate, which acts as an order parameter of the phase transition to the superfluid phase at large chemical potential. However, in the theory with the static quark determinant, the diquark condensate cannot be calculated reliably as its expectation value is always trivial. Therefore, we could not check the validity of the phase diagram within our framework. Nevertheless, our findings hint at the existence of a transition to a superfluid phase.

\subsection{Model and notations} 

Let us begin by establishing some general notations. The $d$-dimensional lattice with $L$ spatial points will be represented by the symbol $\Lambda=Z^d$.
Each single point will be denoted by $x$ with $x\in[0,L-1]$. There are $N_l$ links on the lattice. Given an arbitrary site $x$, $2d$ links are attached to it,
and the one pointing in the direction $n=1,\ldots,d$ will be symbolized by $l=(x,n)$. The unit vector in direction $n$ is $e_n$.

We will define the Polyakov loop $W(x)$ in the standard manner,
\begin{equation}
W(x) \ = \ \mbox{Tr}  \ U(x) \ \ , \ \  
U(x) \ = \ \prod_{t=1}^{N_t} \ U_0(x,t) \ , 
\label{PL_def}
\end{equation}
where $N_t$ is the length of the temporal direction of the $(d+1)$-dimensional space--time lattice where the full theory is usually defined.
The partition function of the $SU(N)$ Polyakov-loop model in the strong coupling approximation and with exact static quark determinant on $\Lambda$ will be called $Z$ and reads
\begin{eqnarray}
		&&Z\equiv Z_{\Lambda}(\beta,m_f,\mu_f; N) =
	[C_0(\beta_{\rm t})]^{N_l} \int \ \prod_x \ dU(x)
	\prod_{l}\left(1+ \lambda \ {\rm Re}W(x) W^{\dagger}(x+e_n) \right) \nonumber \\  
	&& \times \ \prod_x \prod_{f=1}^{N_f} A_f \det  \left[1+ h_+^f U(x) \right]  
	\det \left[1 +  h_-^f U^{\dagger}(x)  \right]  \ . 
	\label{PF_spindef}
\end{eqnarray}
The determinants are taken over group indices and the notation is as follows. For $SU(N)$ the effective coupling constant $\lambda$ is related to the temporal gauge coupling $\beta_{\rm t}$ by 
\begin{eqnarray}
	\lambda \ = \ 2 \left ( \frac{C_{\rm F}(\beta_{\rm t})}{N C_{0}(\beta_{\rm t})}
	\right )^{N_t}
	\ , \ \ \ 
	C_{\rm F}(\beta_{\rm t}) \ = \ \sum_{q=-\infty}^{\infty}
	\ \Bigl.{\rm det} I_{r_i - i + j + q}(\beta_{\rm t})\Bigr|_{1\leq i,j \leq N} \ , 
	\label{D_coeff}
\end{eqnarray}
where $I_n(x)$ is the modified Bessel function and $r_i$ refers to the fundamental representation of $SU(N)$ and is equal to $r_i=\delta_{1i}$.
In this paper we shall use the staggered fermions. Hence, the constants from the quark determinant are given by 
\begin{equation}
	A_f = h_f^{-N } \ , \ \ \ h^f_{\pm} = h_f e^{\pm N_t \mu_f } \ , \ \ \ 
	h_f = e^{-N_t \sinh^{-1} m_f} \ . 
	\label{hpm_stag}
\end{equation}
Here, $m_f=a_t m_f^{\rm ph}$ and $\mu_f=a_t \mu_f^{\rm ph}$ are dimensionless lattice masses and chemical potentials
for each flavor $f$.

\section{Polyakov-loop and Ising models}

The Polyakov-loop model, defined in Eq.~(\ref{PF_spindef}), can be easily mapped onto the two-component Ising model. In order to do this, we use the identity 
\begin{eqnarray}
\label{ising_map}
 &&1 +\frac{\lambda}{2} \, W(x) W^{\dagger}(x+e_n)+ \frac{\lambda}{2} \, W(x)^{\dagger} W(x+e_n)   \nonumber\\ 
 &&=\frac{1}{4} \ \sum_{s, t =\pm 1} \left ( 1+ \sqrt{\frac{\lambda}{2}} \, s W(x) + \sqrt{\frac{\lambda}{2}} \, t W^{\dagger}(x) \right ) \\
&& \times\left ( 1+ \sqrt{\frac{\lambda}{2}} \, s W^{\dagger}(x+e_n) + \sqrt{\frac{\lambda}{2}} \, t W(x+e_n) \right ) .\nonumber 
\end{eqnarray}
Then, the partition function takes the form 
\begin{gather}
\label{PF_ising}
Z = \left [ C_0(\beta_{\rm t}) \right ]^{N_l} \ \frac{1}{4^{N_l}}  \ 
\sum_{\{ s(l), t(l) \} =\pm 1} \prod_x B(x) \ ,  \\ 
\label{Bx_def}
B(x)  \equiv \int \  dU    \prod_{f=1}^{N_f} A_f \det  \left[1+ h_+^f U\right] 
\det \left[1 +  h_-^f U^{\dagger}  \right]  \nonumber \\
\times \prod_{l\in x }  \left ( 1 + \sqrt{\frac{\lambda}{2}} \ s(l) \ W + \sqrt{\frac{\lambda}{2}} \ t(l) \ W^{\dagger} \right ) \ ,
\end{gather}
where the links $l$ are attached to $x$ and $s(l),t(l)$ represent two independent Ising variables attached to those links.

Equivalently, one can write
\begin{gather}
	\label{Bx_def_1dQCS}
B(x)  = Z_0 \ \left \langle 
	 \prod_{l\in x }  \left ( 1 + \sqrt{\frac{\lambda}{2}} \ s(l) \ W + \sqrt{\frac{\lambda}{2}} \ t(l) \ W^{\dagger} \right ) \right \rangle_0 \ ,  
\end{gather}
where $Z_0$ is the partition function of the 1-dimensional QCD,
\begin{gather}
	\label{Z0_def}
	Z_0  = \int \  dU    \prod_{f=1}^{N_f} A_f \det  \left[1+ h_+^f U\right] 
	\det \left[1 +  h_-^f U^{\dagger}  \right]\; ,
\end{gather} 
and the expectation value $\langle \ldots \rangle_0$ refers to $Z_0$. 
Thus, the problem of the determination of effective couplings of the Ising model boils down to the computation of averages of Polyakov-loop powers over one-dimensional QCD, considered first in Ref.~\cite{bilic_88}. 
The resulting representation of the Boltzmann weight is valid for any $SU(N>2)$ model. All effective couplings appearing in it
are strictly positive, hence this formulation is free of sign problems, unlike in Eq.~(\ref{PF_spindef}) where the Boltzmann weight can be negative even in the pure
gauge case. Below we specify this formulation for the $SU(2)$ case.

\subsection{$SU(2)$ with one flavor of staggered fermions} 

For the $SU(2)$ model the above representation simplifies due to the fact that all $SU(2)$ characters are real. Therefore, it is sufficient to introduce only a one-component Ising model.
Moreover, only one flavor of quarks will be introduced and for that reason from here on we drop every flavor index $f$. Precisely, one obtains
\begin{eqnarray} 
\label{PF_su2}
Z &=& \left [ C_0(\beta_{\rm t}) \right ]^{N_l} \ \frac{1}{2^{N_l}}  \ 
\sum_{\{ s(l) \} =\pm 1} \prod_x B(x) \ , \\
B(x)  &=& h^{-2} \int_0^{2 \pi}  \,  d \omega \, \sin^2 \omega
    \left[1+ h_+ e^{i \omega}\right] \left[1 +  h_-  e^{-i \omega}  \right] \left[1+ h_+ e^{- i \omega}\right] \left[1 +  h_-  e^{i \omega}  \right] \nonumber \\ 
 &\times& \prod_{1\in x }  ( 1 + 2 \sqrt{\lambda} \, s(l) \cos \omega ) \ . 
 \label{Bx_su2}
\end{eqnarray}
This expression enables us to write explicitly the Boltzmann weight $B(x)$ of the effective Ising model. In $d$ spatial dimensions this weight is 
\begin{eqnarray} 
\label{Bx_su2_all_dim}
B(x) &=& Z_0 \left [ 1 + \sum_{k=1}^{2d} \ G_k \ \sum_{i_1 < i_2 < \ldots < i_k}^{2d} \ 
s(l_{i_1}) s(l_{i_2}) \ldots s(l_{i_k})   \right ] \ ,  
\end{eqnarray} 
where $G_k$ are the couplings that govern the interaction between Ising spins
\begin{eqnarray}
\label{Gk_coupl_su2}
G_k &=& \left ( 2 \sqrt{\lambda} \right )^k \ 
\left \langle \cos^k \omega \right \rangle_0 \ ,
\end{eqnarray}
and $l_i$ are the $2d$ links attached to the site $x$.

\subsection{Effective couplings of the Ising model} 

In this Subsection we give the explicit expressions for the effective Ising couplings $G_k$. In what follows we use the following notations:
\begin{equation}
\label{ab_def}
a \equiv \cosh(m + \mu) \ \cosh(m-\mu) \ \ , \ \  
b \equiv 2 \cosh m \cosh\mu \ . 
\end{equation}
where
\begin{equation}
\label{---tre}
  m\equiv N_t \sinh^{-1} a_t m^{\rm ph}\ \ , \ \
  \mu\equiv N_t a_t \mu^{\rm ph}\ .
\end{equation}
The partition function $Z_0$ and the expectation values are given by 
\begin{equation}
\label{Z0_su2} 
Z_0 = 1 + 4 a \ , 
\end{equation}
\begin{equation}
\label{cos_exp_su2_evenk} 
\left \langle \cos^{2 k} \omega \right \rangle_0 = \frac{2}{1+4 a} \ 
\frac{\Gamma(k+\frac{1}{2})}{\sqrt{\pi} (k+2)!} \ 
\left ( 1+4 a + 2 k (1+a) \right ) \ , 
\end{equation}
\begin{equation}
\label{cos_exp_su2_oddk} 
\left \langle \cos^{2 k + 1} \omega \right \rangle_0 = 
\frac{2 b}{1+4 a} \ \frac{\Gamma(k+\frac{1}{2})}{\sqrt{\pi} (k+2)!} \ (2k + 1) \ . 
\end{equation}
Collecting all formulas, we find that the Ising couplings~(\ref{Gk_coupl_su2}) for the $SU(2)$ model with fermions are
\begin{eqnarray}
\label{G1_su2_res} 
G_0 = 1 \ , \ G_1 = 2 \ \lambda^{\frac{1}{2}} \ \frac{b}{1+4a}  \ , \  
G_2 = 4 \ \lambda \  \frac{1+2a}{2+8a}   \ , \ 
G_3 = 8 \ \lambda^{\frac{3}{2}} \  \frac{b}{2+8a}  \ ,  \\
\label{G4_su2_res} 
G_4 = 16 \ \lambda^{2} \  \frac{5+8a}{16+64a}  \ , \
G_5 = 32 \ \lambda^{\frac{5}{2}} \  \frac{5 b}{16+64a}  \ ,  \
G_6 = 64 \ \lambda^{3} \  \frac{7+10a}{32+128a}  \ .  \nonumber 
\end{eqnarray}
For the pure gauge model we have that the couplings with odd indices vanish, $G_1=G_3=G_5=0$, while those with even indices read
\begin{eqnarray}
\label{Gk_puresu2_res} 
G_0 = 1 \ , \	G_2 =  \lambda \  ,  \ G_4  = 2  \lambda^2 \ , \ 
G_6 = 5 \ \lambda^3  \ .
\end{eqnarray} 
We will present the necessary observables in the next Section in terms of the shifted couplings 
\begin{eqnarray}
\label{Gk_shifted_su2_res} 
G_{k,i} = 2^k \ \lambda^{\frac{k}{2}} \  \left \langle \cos^{k+i} \omega \right \rangle_0 \ . 
\end{eqnarray}
{\it E.g.}, the shifted couplings for the pure gauge model read
\begin{eqnarray}
\label{Gki_pure gauge} 
&&G_{2k,1} = G_{2k+1,2} = 0 \ , \\   	
&&G_{2k+1,1} = 2^{2k+1} \ \lambda^{\frac{(2k+1)}{2}} \ 
\frac{\Gamma(k+3/2)}{\sqrt{\pi} (k+2)!} \  , \ 
G_{2k,2} = 2^{2k} \ \lambda^{k} \ \frac{\Gamma(k+3/2)}{\sqrt{\pi} (k+2)!}  \  . 
\end{eqnarray}
In the large $N_t$ and $\beta_t={a_s}/({a_tg^2})$ limits ({\it i.e.} the finite temperature limit) the pure gauge coupling constant 
\begin{equation}
\label{lambda_su2} 
\lambda = \left ( \frac{I_2(2 \beta_{\rm t})}{I_1(2 \beta_{\rm t})} \right )^{N_t}
\end{equation}
becomes 
\begin{equation}
\label{lambda_su2_largeNt} 
\lambda \approx  \exp \left [ - \frac{3 N_t}{4 \beta_t} \right ] = 
\exp \left [ - \frac{3 g^2}{4 a_s T} \right ] \ . 
\end{equation}

\subsection{Exact solution of the 1-dimensional model}
\label{1d_su2}

\begin{figure}[htb]
\centering
\includegraphics[width=0.48\linewidth,align=t,clip]{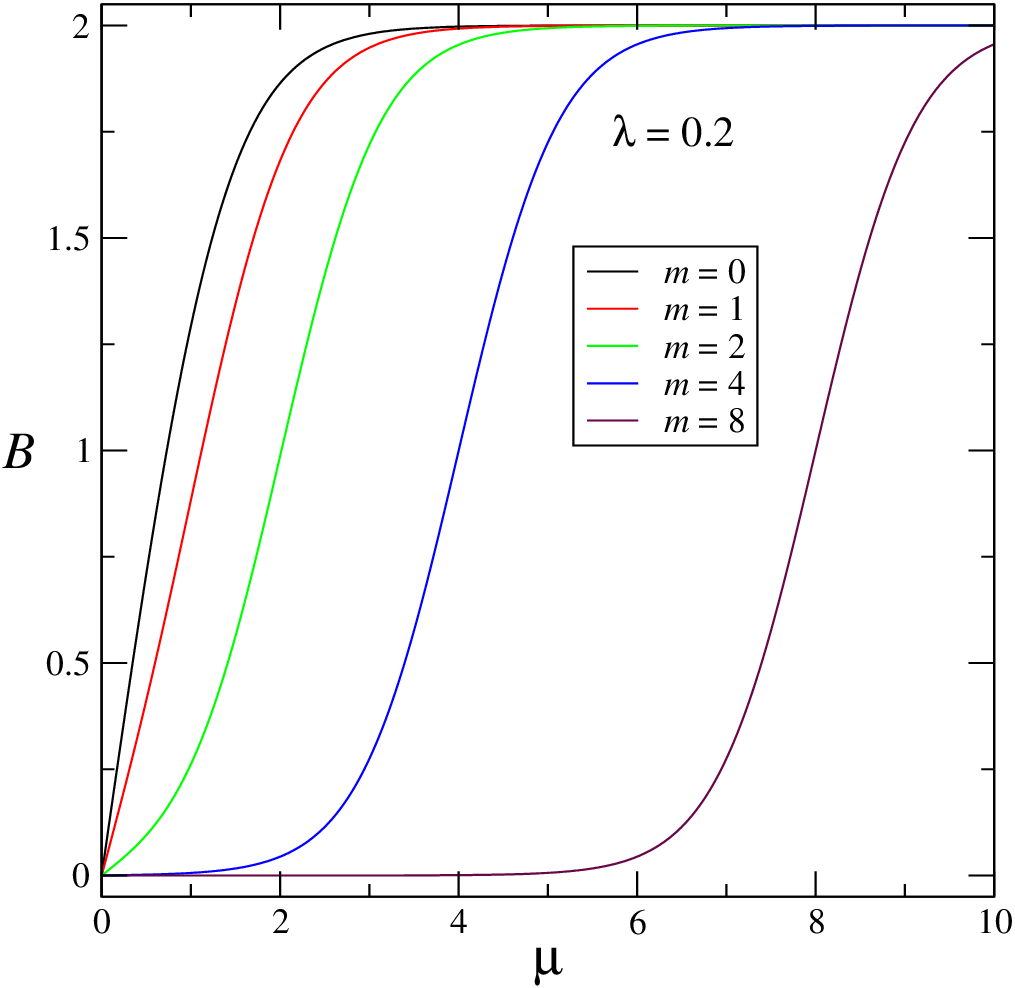}
\hspace{0.3cm}
\includegraphics[width=0.48\linewidth,align=t,clip]{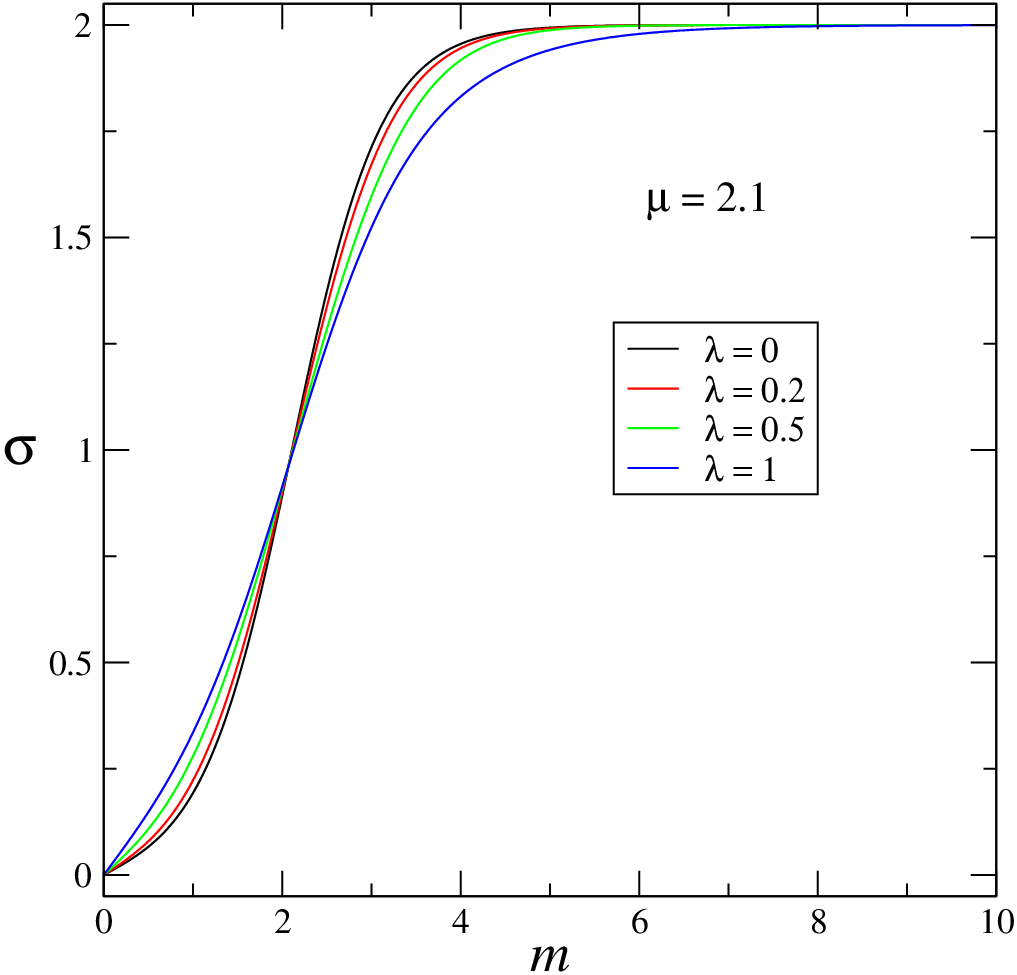}
\caption{(Left) Quark density {\it versus} chemical potential with  the coupling $\lambda=0.2$, $m=0,1,2,4,8$. (Right) Quark condensate {\it versus} $m$, for  $\mu=2.1$, $\lambda=0,0.2,0.5,1$.}
\label{fig:1d-BD-QC}
\end{figure}

As the simplest example, let us study the above model in one dimension. It can be solved exactly, obtaining for the partition function
\begin{eqnarray} 
\label{PF_su2_1d}
Z_{1d} = \left [ C_0(\beta_{\rm t}) \right ]^{L N_t} \  \sum_i \Lambda_i^L  \ ,
 \label{ggggg2}
\end{eqnarray}
where $\Lambda_i$ are the eigenvalues of the transfer matrix constructed from~(\ref{Bx_su2}),
\begin{equation}
\label{transfer_matr_1d}
T_{m,n}= \left(
\begin{array}{cc}
 2 \lambda+1+ 4  (\lambda+1)a+4 \sqrt{\lambda} b  & 1-2 \lambda-4 (\lambda-1) a \\
 1-2 \lambda-4 (\lambda-1)a & 2 \lambda+1 +4 (\lambda+1)a -4 \sqrt{\lambda} b\\
\end{array}
\right) \ , 
\end{equation} 
\begin{equation} 
\label{eigenvalues_1dsu2}
\Lambda_{1,2} = 4(\lambda+1) a + 2 \lambda + 1 \pm 
\sqrt{ (4 (\lambda-1) a + 2 \lambda-1)^2+16 \lambda b^2} \ . 
\end{equation}
As expected, there is no phase transition in 1-dimension, even in the pure gauge theory. 
In Fig.~\ref{fig:1d-BD-QC} we show the quark density and the quark condensate for various values of the parameters and
couplings in the thermodynamic limit $L\to\infty$ (only the largest eigenvalue $\Lambda_1$ contributes to this limit). 

\begin{figure}[t]
\centering
\includegraphics[width=0.48\linewidth,align=t,clip]{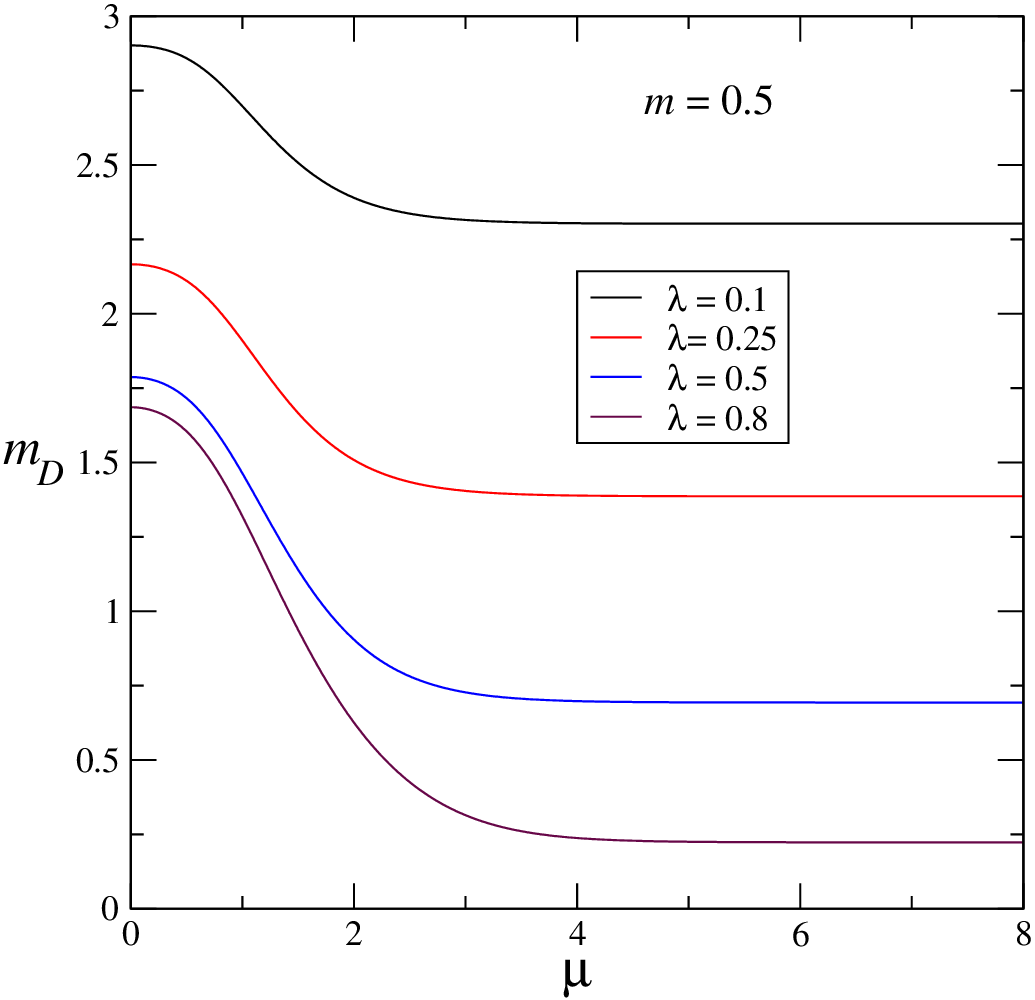}
\hspace{0.3cm} 
\includegraphics[width=0.48\linewidth,align=t,clip]{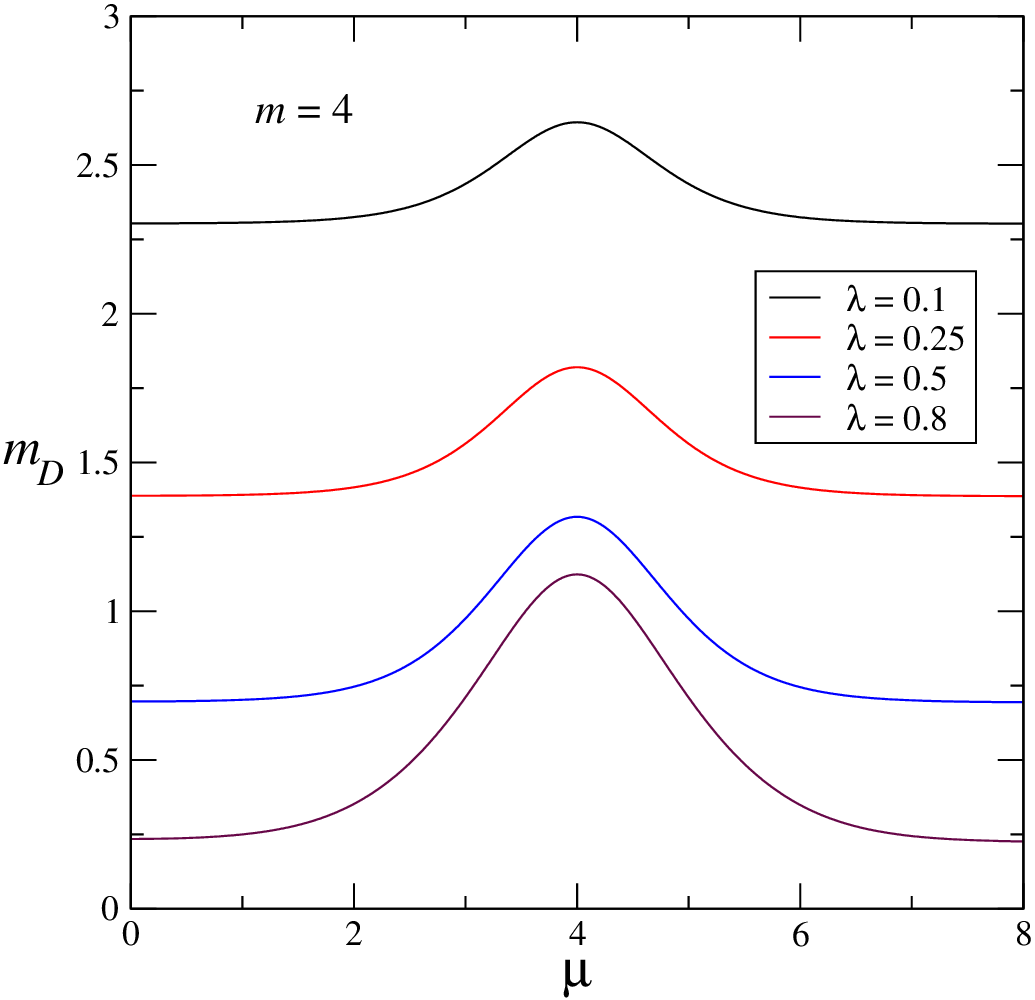}
\caption{Screening masses {\it versus} chemical potential for  $m=0.5$ (left),  
$m=4$ (right) and the values $\lambda=0.1,0.25,0.5,0.8$ of the coupling $\lambda$.}
\label{fig:1d_screen_mass}
\end{figure}

It is also instructive to look at the screening mass as a function of the chemical potential. This mass is given by 
\begin{equation}
\label{1d_screen_mass}
m_D = - \ln \left ( \frac{\Lambda_2}{\Lambda_1}  \right )  
\end{equation}
and it is plotted in Fig.~\ref{fig:1d_screen_mass}. At small values of quark masses, $m\lesssim 0.5$, the maximum of the screening mass is found at $\mu=0$. 
For all masses $m\gtrsim 1$, the maximum is reached for $m=\mu$. This is probably related to a qualitatively different behavior of the quark density at small $m$ and large $m$, as it can be seen from the left panel of Fig.~\ref{fig:1d-BD-QC}.
A similar behavior of the screening mass remains valid for the 3-dimensional model, as we shall demonstrate by numerical simulations in Section~\ref{full_theory}.

\section{Observables and Monte-Carlo update} 

In this Section we introduce the observables that interest us and describe the algorithm utilized for Monte-Carlo updating
the effective Ising model in $d=3$ dimensions.

\subsection{Observables}
\label{SubsecObservables}

To begin with, we introduce the following notation:
\begin{equation}
	H_i(x) = \sum_{k=0}^{2d} \ G_{k,i} \ \sum_{i_1 < i_2 < \ldots < i_k}^{2d} \ 
	s(l_{i_1}) s(l_{i_2}) \ldots s(l_{i_k})	\ .
	\label{Hki_def}
\end{equation}
This notation will be practical in that most of the expectation values of the observables listed below can be cast in terms of~(\ref{Hki_def}).
For example, all considerations refer to the partition function
\begin{eqnarray} 
	\label{PF_su2_ising} 
	Z = \frac{1}{2^{N_l}}  \ \sum_{\{ s(l) \} =\pm 1} \prod_x \ H_0(x)  \  . 
\end{eqnarray}
The observables that we will measure in the present study are:

1. Polyakov loop and its susceptibility
\begin{eqnarray}
	\label{su2_PL} 
	&&\frac{1}{2} \ \langle W(x)  \rangle =  \frac{1}{L^d} \ \sum_x \ 
	\left \langle \frac{H_1(x)}{H_0(x)}  \right \rangle  \ ,    \\ 
	\label{su2_PL_susc} 
	&&\chi_W = \frac{1}{4} \ \langle W(x)^2  \rangle - \frac{1}{4} \ \langle W(x)  \rangle^2 \\ 
	&&=  \frac{1}{L^d} \left [ \sum_x \ \left \langle \frac{H_2(x)}{H_0(x)}  \right \rangle 
	+ \sum_{x\ne y} \ \left \langle \frac{H_1(x)}{H_0(x)} \frac{H_1(y)}{H_0(y)} \right \rangle 
	- \left ( \sum_x \ 
	\left \langle \frac{H_1(x)}{H_0(x)}  \right \rangle \right )^2 \right ] \ .  \nonumber 
\end{eqnarray}

2. Correlation function of Polyakov loops 
\begin{eqnarray}
	\label{su2_PL_corr} 
	\frac{1}{4} \ \langle W(x) W(y)  \rangle =  
	\left \langle \frac{H_1(x)}{H_0(x)} \frac{H_1(y)}{H_0(y)} \right \rangle  \  .   
\end{eqnarray} 

3. Quark density 
\begin{eqnarray}
	\label{su2_density}
	B = \frac{1}{L^d} \ \frac{\partial \ln Z}{\partial \mu} = 
	\frac{4\sinh 2\mu}{1+4a} + \frac{1}{L^d} \ \sum_x \
	\left \langle \frac{\partial_{\mu} H_0(x)}{H_0(x)}  \right  \rangle \ .
\end{eqnarray}

4. Quark condensate 
\begin{eqnarray}
	\label{su2_condensate}
	\sigma = \frac{1}{L^d} \ \frac{\partial \ln Z}{\partial m} = 
	\frac{4\sinh 2m}{1+4a} + \frac{1}{L^d} \ \sum_x \
	\left \langle \frac{\partial_{m} H_0(x)}{H_0(x)}  \right  \rangle \ .
\end{eqnarray}

5. Susceptibility and Binder cumulant for the Ising spin 
\begin{eqnarray}
  \overline{s}\equiv\frac{1}{N_l}\sum_l s(l)\ , &&
  \label{su2_Ising_susc}
	\chi = L^d \ \left( \langle \overline{s}^{\,2} \rangle - \langle |\overline{s}| \rangle^2 \right) \;, \\
	\label{su2_Ising_Binder}
	B_4^{(s)}& =& 1-\frac{ \langle \overline{s}^{\,4} \rangle}{3\langle \overline{s}^{\,2} \rangle^2} \;.
\end{eqnarray}

\subsection{Monte-Carlo method}

The partial integrations described in the previous Section yield a simplified version of the Polyakov-loop model, which is particularly suitable for numerical simulations.
The input parameters for a Monte-Carlo simulation are, apart from the lattice temporal and spatial sizes, the value of $\lambda$ in~(\ref{D_coeff}), and the quark mass $m$ and chemical potential $\mu$.
Moreover, the dynamical variables of the model are reduced to a set of Ising spins $s(l)=\pm1$, one per link $l$. However, as~(\ref{PF_ising}) and~(\ref{Bx_su2_all_dim}) show,
the interaction between the several Ising variables takes place at the sites, not on the links. Therefore, our Monte-Carlo method consists in the following steps: {\it (i)} all links are visited;
{\it (ii)} given one link $l$, its two endpoints $x_1$ and $x_2$ are individuated, and {\it (iii)} the contribution to~(\ref{PF_ising})
\begin{equation}
  B(x_1)B(x_2)\;,
\end{equation}
is calculated, first with the current value of the variable $s(l)$, and then with the flipped value $-s(l)$. Finally, the ratio of those two results is submitted to a Metropolis
test~\cite{Metropolis} in order to decide which of the two values, $s(l)$ or $-s(l)$, is dynamically preferred. There exist other simulation techniques which are usually
more efficient than Metropolis, like global cluster algorithms or local algorithms based on the possibility to analytically invert the functional form of the action like
Heat-Bath. However, all those possibilities were discarded due to the complicated mathematical dependence of~(\ref{Bx_su2_all_dim}) on the Ising spins $s(l)$.

A naive extension of the above Metropolis-based updating algorithm for pairs of Ising variables is certainly feasible in 
the more interesting case of the theory invariant under the $SU(3)$ gauge group.

\section{Results} 

In this Section we present a detailed account of the results obtained by measuring the observables listed in Section~\ref{SubsecObservables}.
The analyses for the pure gauge and for the full theory cases are described in two separate Subsections.

The numerical Monte-Carlo simulations were performed on lattices with spatial extent $L$ up to $32$. Measurements were taken every 10
lattice updates until collecting in all typically 100k measurements per simulation. The error analysis was based on  a jackknife method applied
to various blocking levels, their bin size varying from 50 to 10k.

\subsection{Critical behavior of the pure gauge model} 

In the first part of this Subsection we describe the results obtained from simulations at finite temporal size $N_t$.

The main aim of our numerical investigation of the pure gauge model was to check by finite-size scaling (FSS)
analysis that it belongs to the same universality class of the underlying $SU(2)$ (3+1)-dimensional lattice gauge theory at
finite temperature, {\it i.e.} the 3-dimensional Ising class, whose critical indices have been determined with
high accuracy in Ref.~\cite{Kos:2016ysd}. In this respect, the observables which turned out to be most convenient
are the absolute value of the Ising link and the related susceptibility $\chi$ defined in~(\ref{su2_Ising_susc}), along with the Binder cumulant $B_4^{(s)}$ defined in~(\ref{su2_Ising_Binder}).

\begin{figure}[htb]
\centering
\includegraphics[width=0.47\linewidth,clip]{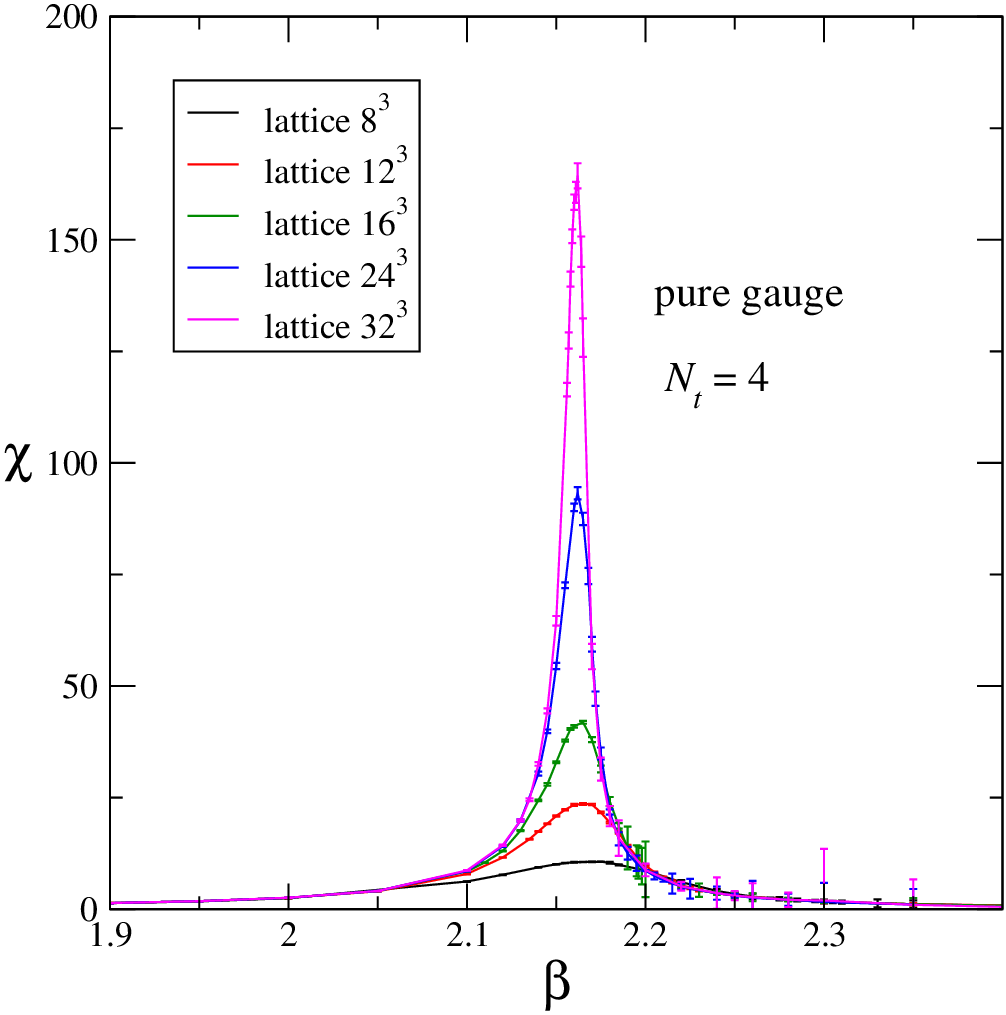}
\caption{Susceptibility of the absolute value of the Ising link variable {\it versus} the coupling $\beta$, in the
  pure gauge theory with $N_t=4$, on various lattice spatial sizes. Solid lines are drawn to guide the eye.}
\label{fig:gauge_susc_Nt=4}
\end{figure}

We considered first the dependence of the Ising link susceptibility on the coupling $\beta$ in the model
with $N_t$=4, on various lattice sizes $L^3$. Results are summarized in Fig.~\ref{fig:gauge_susc_Nt=4} and show that the
height of the peak of the susceptibility increases with the volume, while its position moves slowly to the left, taking
a value about~2.16 on the largest lattice. This value is comparable to the value 2.29895(10) obtained in
Ref.~\cite{Engels:1995em} for the $SU(2)$ (3+1)-dimensional lattice gauge theory at finite temperature in the thermodynamic limit.
 
The rest of this Subsection is devoted to the study of the $N_t=\infty$ case, for which a more systematic FSS analysis has been performed.

\begin{figure}[htb]
\centering
\includegraphics[width=0.47\linewidth,align=t,clip]{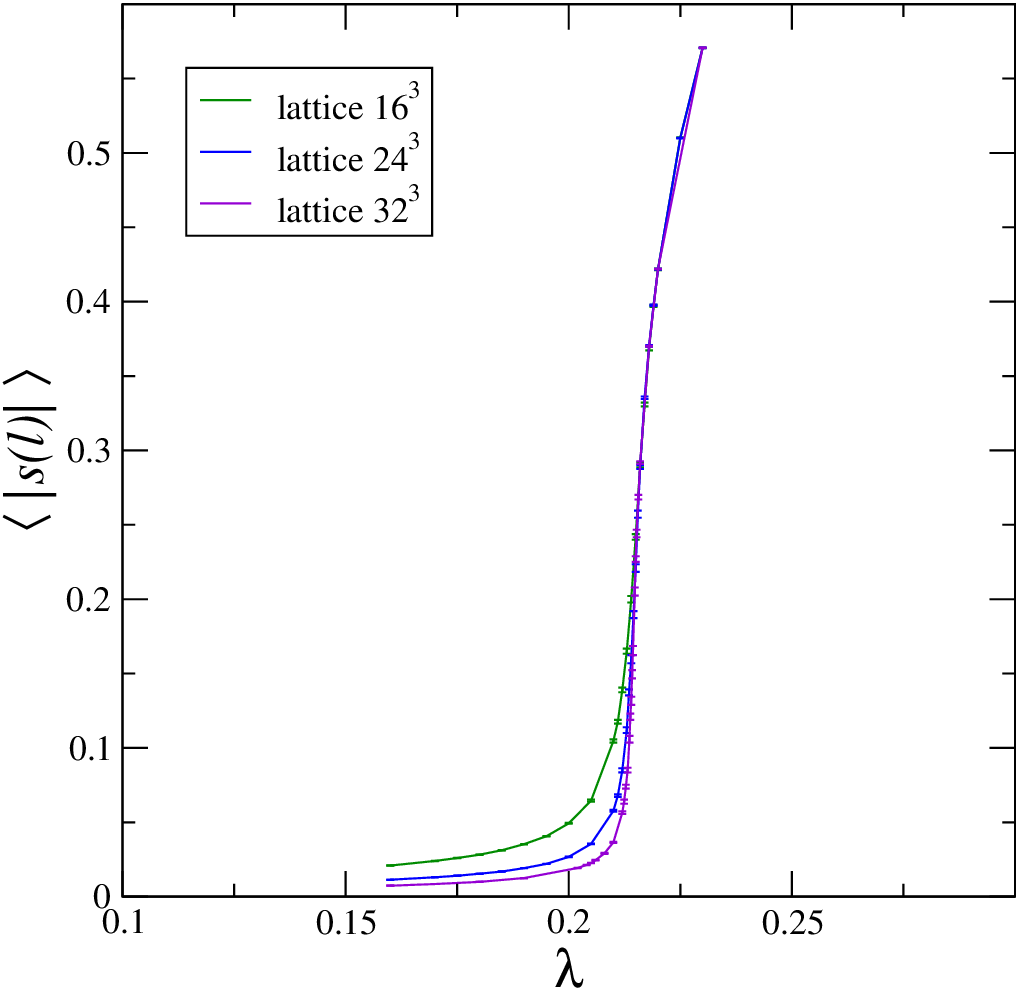}
\hspace{0.3cm}
\includegraphics[width=0.47\linewidth,align=t,clip]{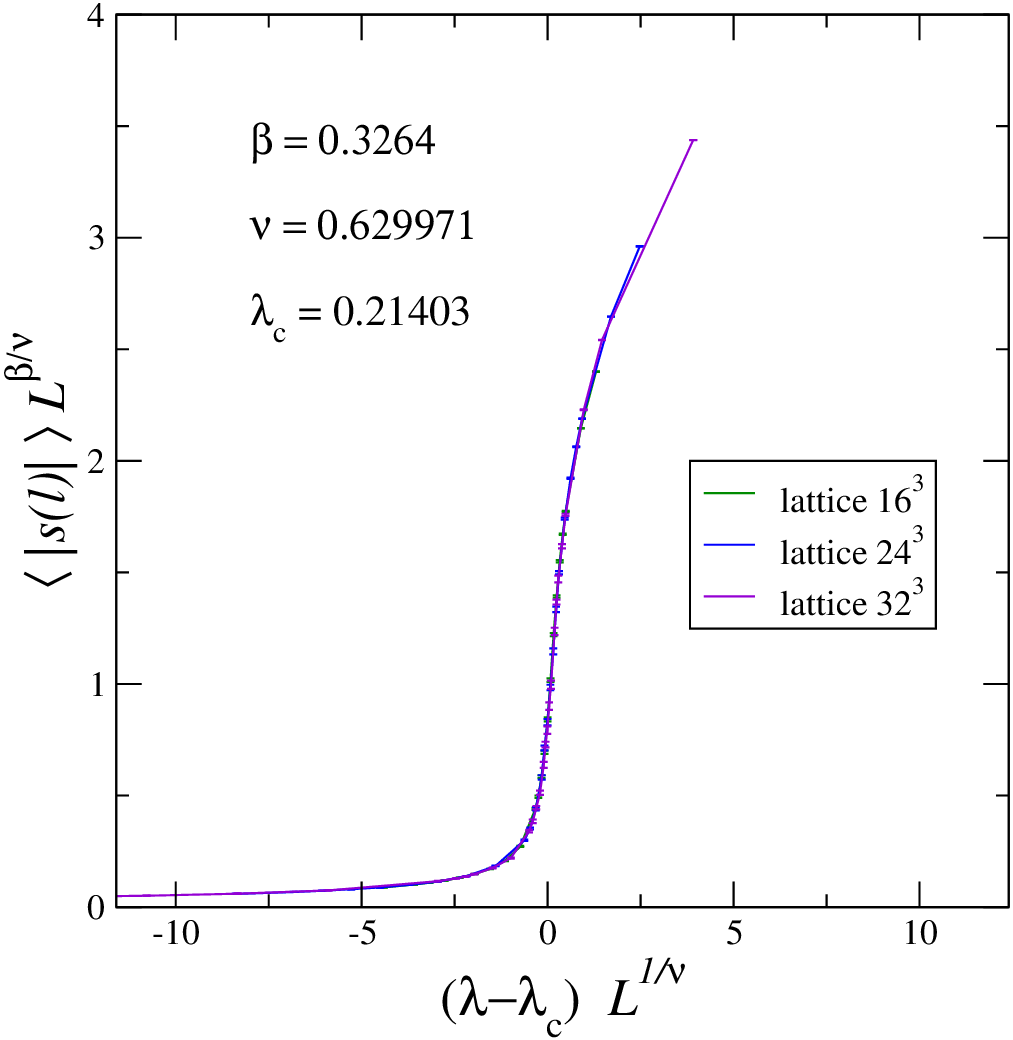}
\caption{(Left) Absolute value of the Ising link variable {\it versus} the coupling $\lambda$, in the
  pure gauge theory at infinite $N_t$, on various lattice spatial sizes. (Right) Same as left, after rescaling.
  Solid lines are drawn to guide the eye.}
\label{fig:gauge_Ising}
\end{figure}

The absolute value of the configuration-averaged Ising link, shown in Fig.~\ref{fig:gauge_Ising}(left), exhibits a transition from the (confined) low-$\lambda$ region,
where it is nearly zero, to the (deconfined) high-$\lambda$ region, the jump getting steeper as the lattice size increases. After 
rescaling the observable by the factor $L^{\beta/\nu}$ and replacing the coupling $\lambda$ with $(\lambda-\lambda_{\rm c}) L^{1/\nu}$,
for indices $\beta$ and $\nu$ taken from literature, and using our determination of the critical coupling $\lambda_{\rm c}$ (see below),
data points obtained on different lattice sizes nicely collapse on a universal curve, as shown in Fig.~\ref{fig:gauge_Ising}(right).

\begin{figure}[htb]
\centering
\includegraphics[width=0.47\linewidth,align=t,clip]{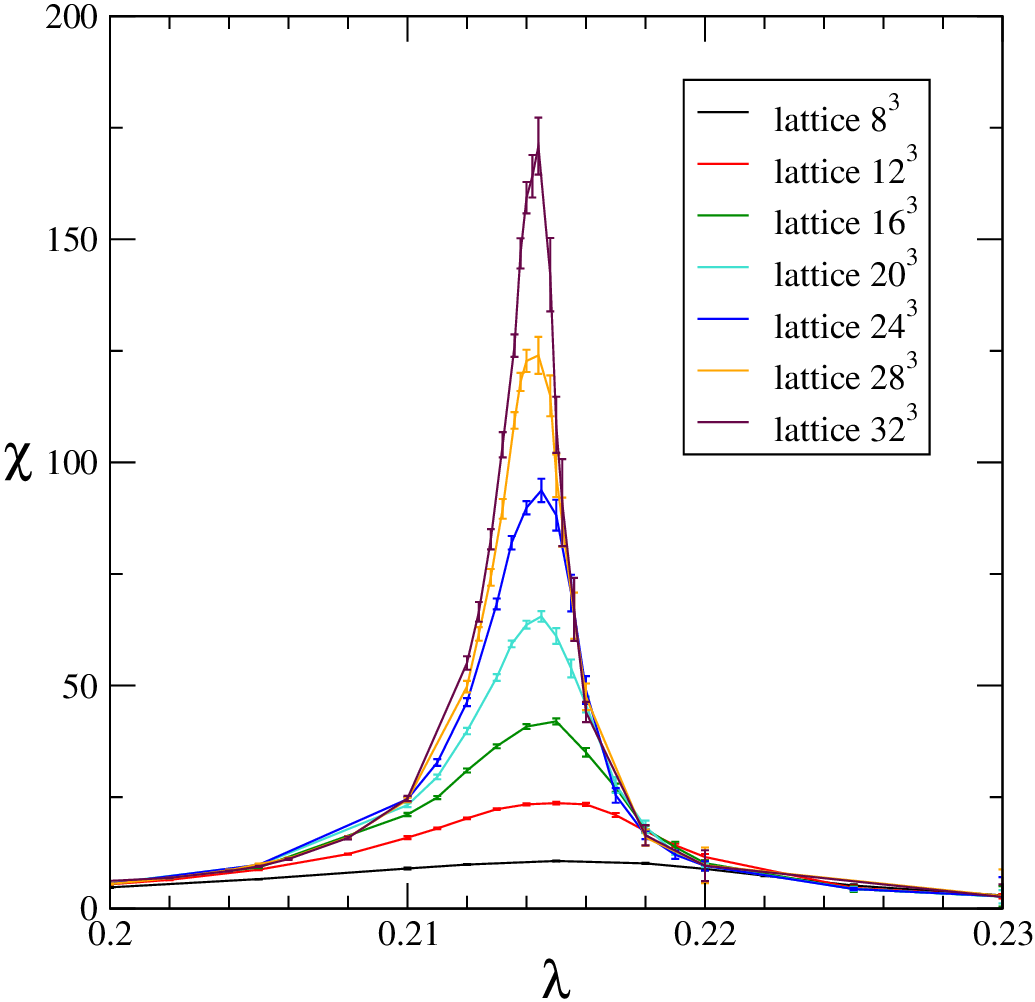}
\hspace{0.3cm}
\includegraphics[width=0.47\linewidth,align=t,clip]{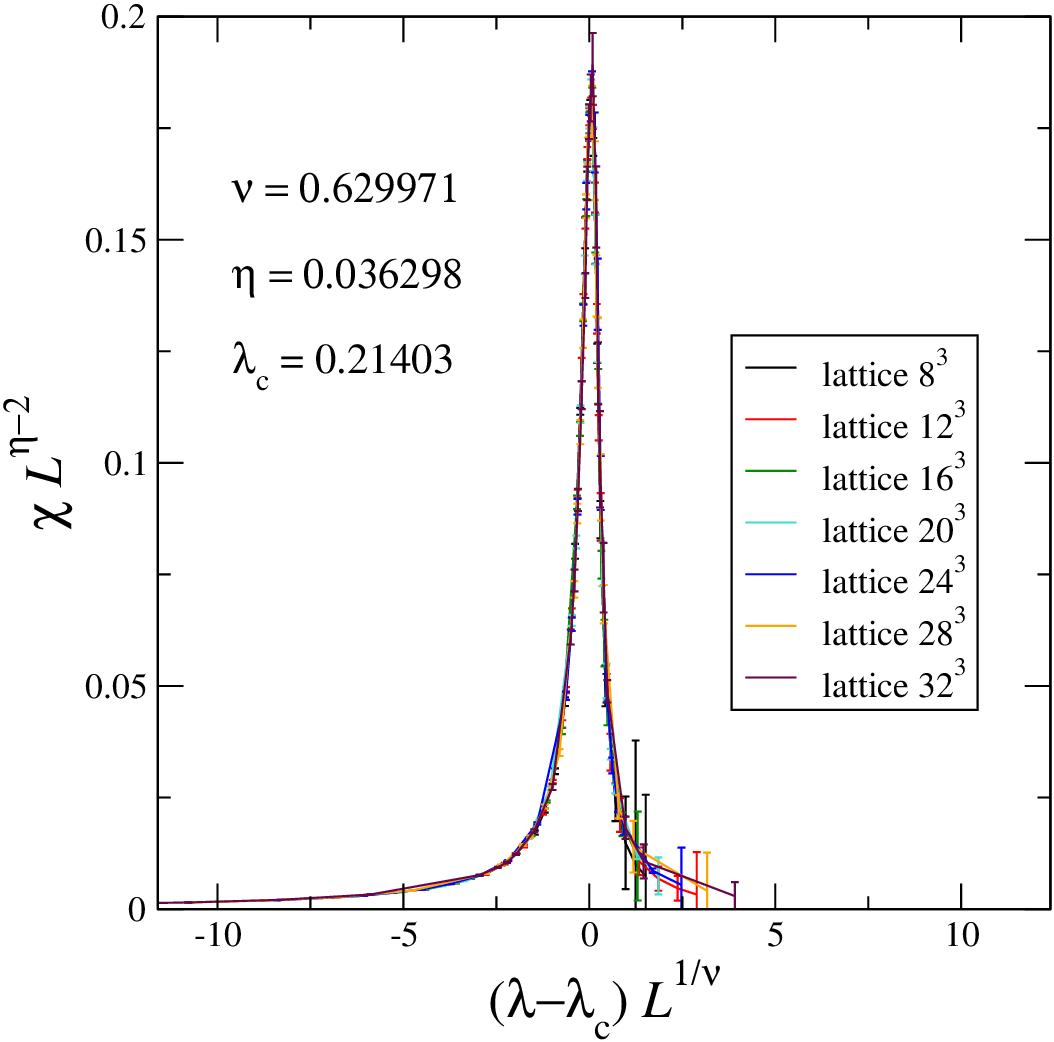}
\caption{(Left) Susceptibility of the absolute value of the Ising link variable {\it versus} the coupling $\lambda$, in the
  pure gauge theory at infinite $N_t$, on various lattice spatial sizes. (Right) Same as left, after rescaling.
  Solid lines are drawn to guide the eye.}
\label{fig:gauge_susc}
\end{figure}

The behavior in $\lambda$ of the susceptibility exhibits the typical peak, which becomes more pronounced as the
lattice spatial size increases, its position slowly moving to the left, see Fig.~\ref{fig:gauge_susc}(left). Also in this case, the standard FSS
rescaling of the observable and coupling leads to a nice collapse, as shown in Fig.~\ref{fig:gauge_susc}(right). The
positions of the maxima of the susceptibility on the different lattice spatial sizes, {\it i.e.} the pseudo-critical coupling $\lambda_{\rm pc}$,
were determined by a Lorentzian fit in the respective peak region and are summarized in Table~\ref{tab:peaks}.

\begin{table}[tb]
  \setlength{\tabcolsep}{10pt}
  \centering
  \caption{Position of the maximum of the Ising link susceptibility $\chi$ on the lattice $L^3$, as determined from a Lorentzian fit near the peak.}
  \vspace{0.2cm}
  \begin{tabular}{rl}
    \hline
    $L$ & \ \ \ $\lambda_{\rm pc}$ \\ 
    \hline
     8  &  0.21514(30)  \\
    12  &  0.21476(26)  \\
    16  &  0.21464(20)  \\ 
    20  &  0.214289(91) \\
    24  &  0.21437(17)  \\
    28  &  0.214152(67)  \\
    32  &  0.214229(58)  \\
    \hline
  \end{tabular}
  \label{tab:peaks}
\end{table}

The values of the pseudo-critical couplings can be used to estimate the critical coupling $\lambda_{\rm c}$ in the thermodynamic
limit by a fit with the scaling function
\[
\lambda_{\rm pc} = \lambda_{\rm c} + \frac{C}{L^{1/\nu}},
\]
which gives $\lambda_{\rm c}$ = 0.21403(73), $C$=0.02(14) and $\nu=0.69(1.34)$, with $\chi^2$/d.o.f.=0.81 and the uncertainties 
on the fit parameters fixed by requiring a 95\% confidence level (this setting applies also to all subsequent fits).

\begin{figure}[htb]
\centering
\includegraphics[width=0.47\linewidth,align=t,clip]{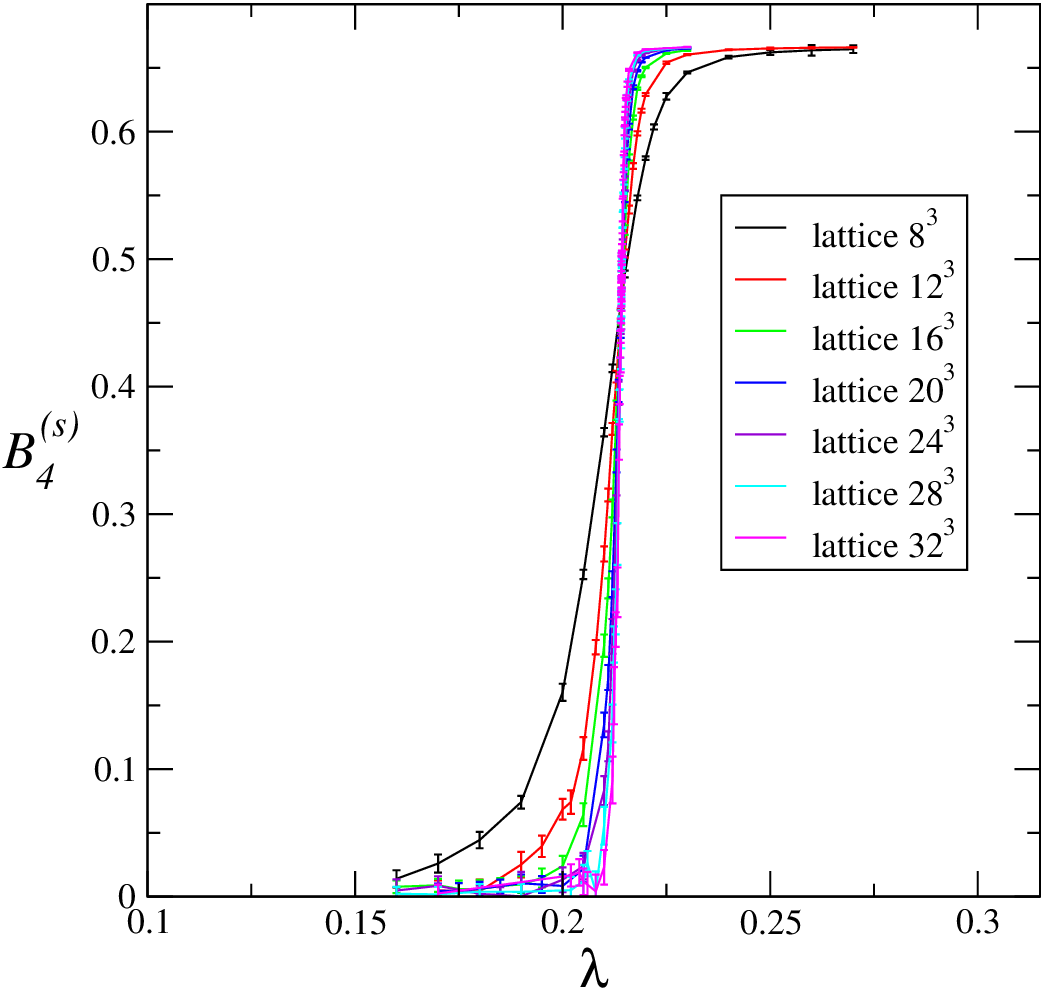}
\hspace{0.3cm}
\includegraphics[width=0.47\linewidth,align=t,clip]{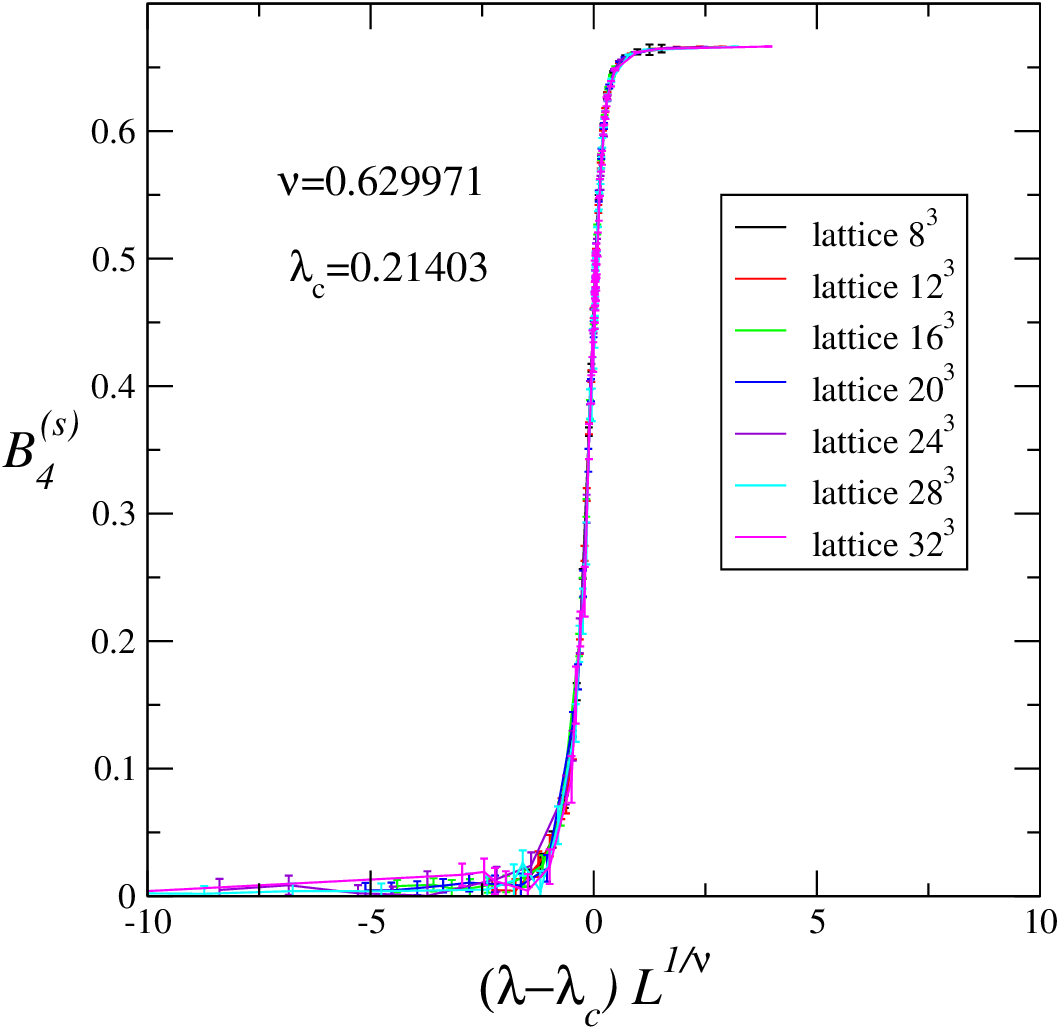}
\caption{(Left) Binder cumulant $B_4^{(s)}$ of the Ising link variable {\it versus} the coupling $\lambda$, in the
  pure gauge theory at infinite $N_t$, on various lattice spatial sizes. (Right) Same as left, after rescaling.
  Solid lines are drawn to guide the eye.}
\label{fig:gauge_Binder}
\end{figure}

The Binder cumulant $B_4^{(s)}$ of the Ising link variable {\it versus} $\lambda$ in the transition region is shown in
Fig.~\ref{fig:gauge_Binder}(left). After standard FSS rescaling, a nice collapse plot is obtained, see Fig.~\ref{fig:gauge_Binder}(right).
The values $B_4^{(s)}$ on a lattice with a given spatial extent $L$ can be used to provide an alternative estimation of $\lambda_{\rm c}$, by a fit with the function
\begin{equation}
B_4^{(s)}(\lambda) = C_1 + C_2 \ (\lambda-\lambda_{\rm c}) \ L^{1/\nu}\;.
\label{fit_Binder}
\end{equation}
For the two largest lattices, the result of the fit is summarized in Table~\ref{tab:fit_Binder}. Unfortunately, we were not able to
obtain a good simultaneous fit of data on $L=32$ and $L=28$.

\begin{table}[tb]
  \setlength{\tabcolsep}{10pt}
  \centering
  \caption{Result of the fit to the $B_4^{(s)}(\lambda)$ data on the lattices $32^3$ and $28^3$ with the function given in~(\ref{fit_Binder}).}
  \vspace{0.2cm}
  \begin{tabular}{cccccc}
    \hline
    $L$ & $C_1$     & $C_2$      & $\lambda_{\rm c}$ & $\nu$      & $\chi^2$/d.o.f \\
    \hline
    32  & 0.420(57) & 0.70(33)   & 0.21383(31) &  0.632(57) & 0.97          \\
    28  & 0.528(64) & 0.90(1.40) & 0.21450(97) &  0.64(24)  & 0.89          \\
    \hline
  \end{tabular}
  \label{tab:fit_Binder}
\end{table}
  
\begin{figure}[htb]
\centering
\includegraphics[width=0.47\linewidth,align=t,clip]{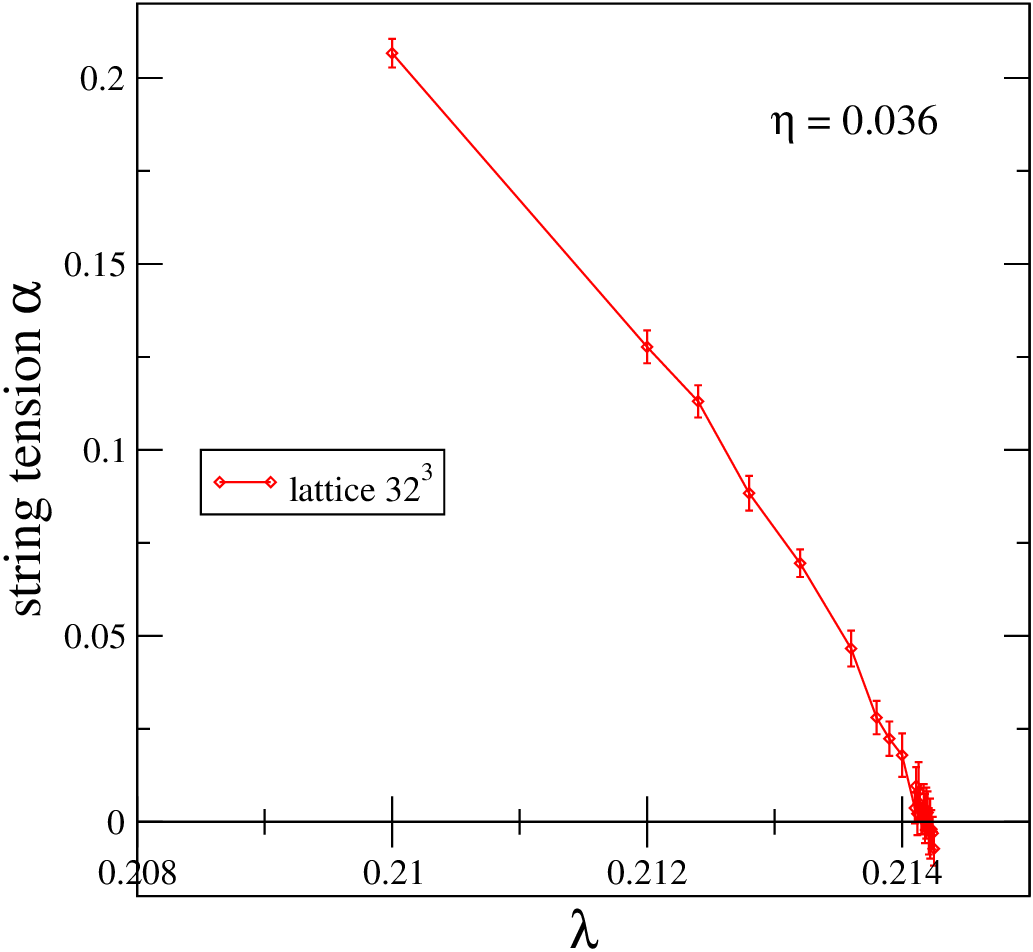}
\caption{String tension {\it versus} the coupling $\lambda$, in the pure gauge theory at infinite $N_t$ on the lattice 32$^3$,
  as determined from a fit to correlation data as explained in the text, with the exponent $\eta$ fixed at 0.036.
  The solid line is drawn to guide the eye.}
\label{fig:gauge_string_tension}
\end{figure}

To conclude, we obtained data for the string tension $\alpha$ below the transition point (they are shown in Fig.~\ref{fig:gauge_string_tension})
and determined their dependence on lambda by fitting the
correlation data of the Polyakov loop, defined in~(\ref{su2_PL_corr}), with the function
\[
C(r) = A \ \left( \frac{\exp(-\alpha r)}{r^{1+c}} + \frac{\exp(-\alpha (L-r))}{(L-r)^{1+c}} \right) \;,\;\;\;\;\;
r=x_1-y_1\;,
\]
with the exponent $c$ fixed to the value of the critical index $\eta \simeq 0.036$. The string tension drops to zero when the critical
$\lambda$-region is approached from below. A fit to the string tension values with the function
\[
\alpha=B \ (\lambda-\lambda_{\rm c})^\nu
\]
gives
\[
B= 10.5(4.1) \;, \;\;\; \lambda_{\rm c} =   0.214087(53)\;, \;\;\; \nu = 0.714(67)\;, \;\;\; \chi^2/{\rm d.o.f} =    0.27.
\]

\subsection{Observables in the full theory} 
\label{full_theory}

In the theory with fermions other two parameters enter the game and they are the fermion mass $m$, and the baryon chemical potential $\mu$.
All results below are presented in terms of the dimensionless quantities $\mu$ and $m$
defined in (\ref{---tre}). The logarithmic derivatives of the partition function with respect to $m$ and $\mu$ define new interesting observables, correspondingly the quark condensate $\sigma$ and the quark density $B$.

\begin{figure}[htb]
\centering
\includegraphics[width=0.47\linewidth,align=t,clip]{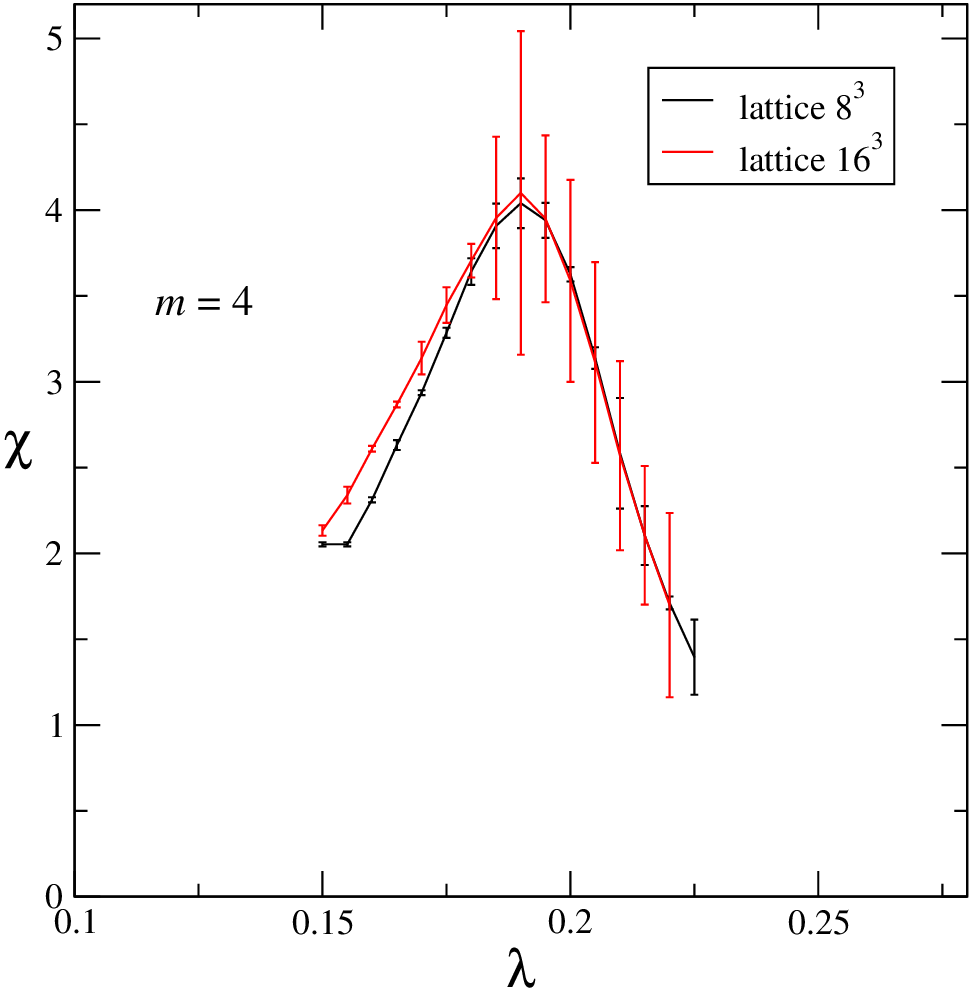}
\caption{Susceptibility of the absolute value of the Ising link variable {\it versus} the coupling $\lambda$, in the
  full theory at infinite $N_t$ and fermion mass $m$=4, on two lattice spatial sizes. Solid lines are drawn to guide the eye.}
\label{fig:full_susc}
\end{figure}

The pure gauge case can be recovered as the infinite mass limit of the theory with fermions, while for finite values of the
mass $m$ we expect a weakening of the transition from the low- to the high-$\lambda$ region, an effect which becomes more visible as
the mass is lowered. For an evidence of that behavior, we calculated the Ising link susceptibility in a wide range of $\lambda$ values for
$m$=4 on two lattice sizes, finding that, differently from the pure gauge case, the height of the susceptibility does not change
with the volume in an appreciable way, as shown in Fig.~\ref{fig:full_susc}.

\begin{figure}[htb]
\centering
\includegraphics[width=0.47\linewidth,align=t,clip]{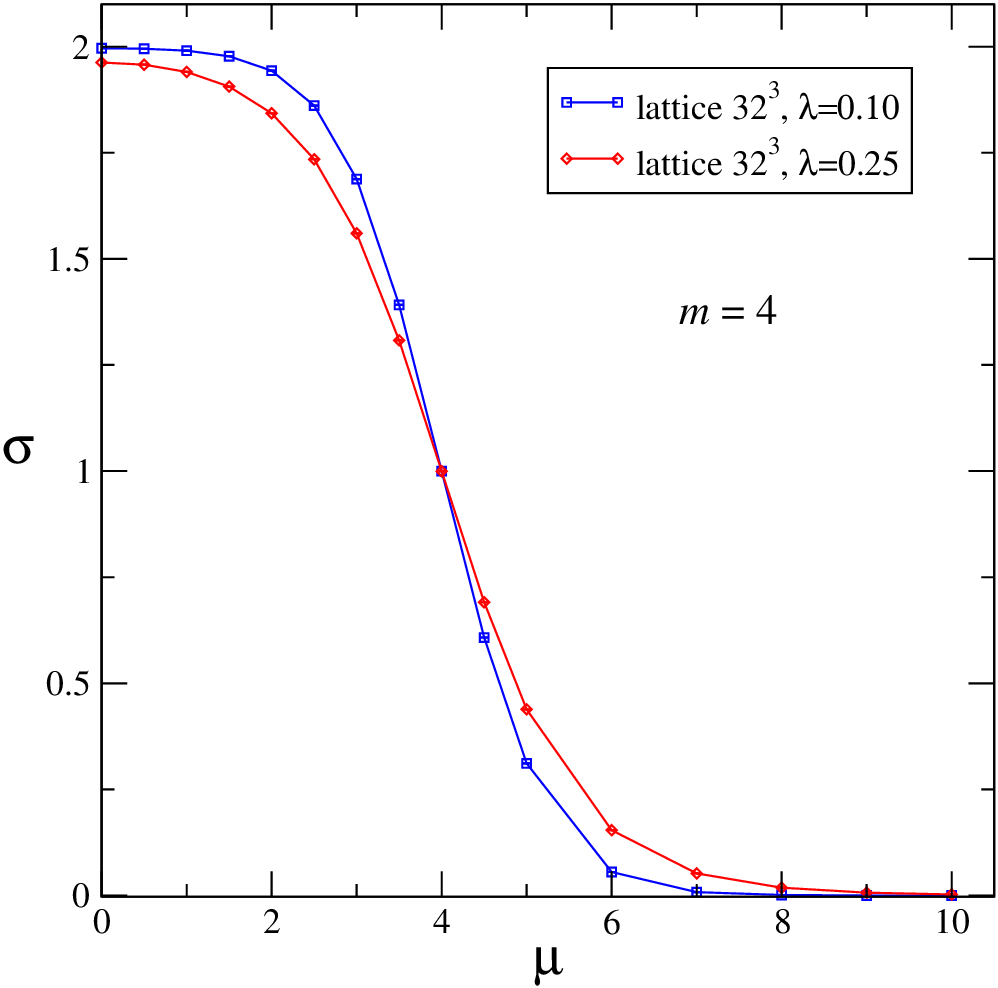}
\hspace{0.3cm}
\includegraphics[width=0.47\linewidth,align=t,clip]{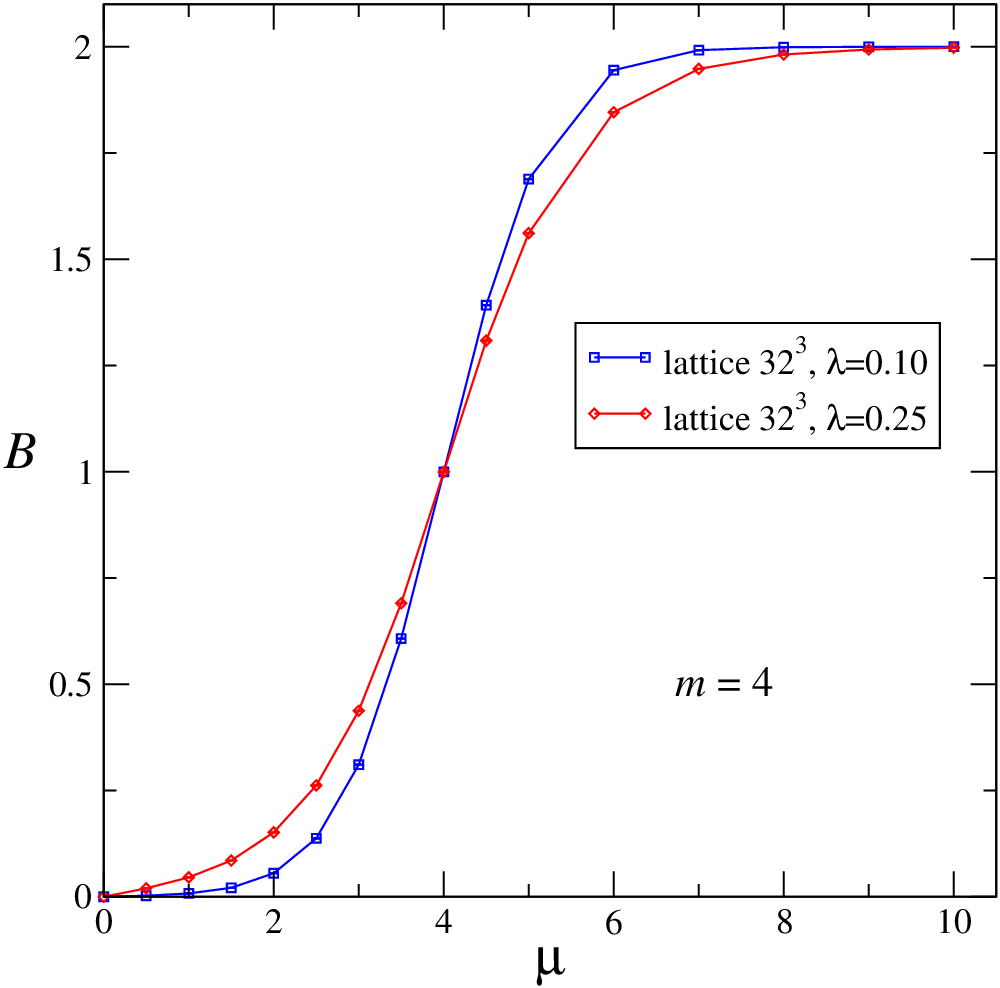}
\caption{(Left) Quark condensate $\sigma$ {\it versus} the chemical potential $\mu$, in the full theory at infinite $N_t$ and
  fermionic mass $m$=4, on lattice 32$^3$ for two values of the coupling $\lambda$. (Right) Same as left, for the quark
  density $B$.
  Solid lines are drawn to guide the eye.}
\label{fig:full_cond_dens_m4}
\end{figure}

\begin{figure}[htb]
\centering
\includegraphics[width=0.47\linewidth,align=t,clip]{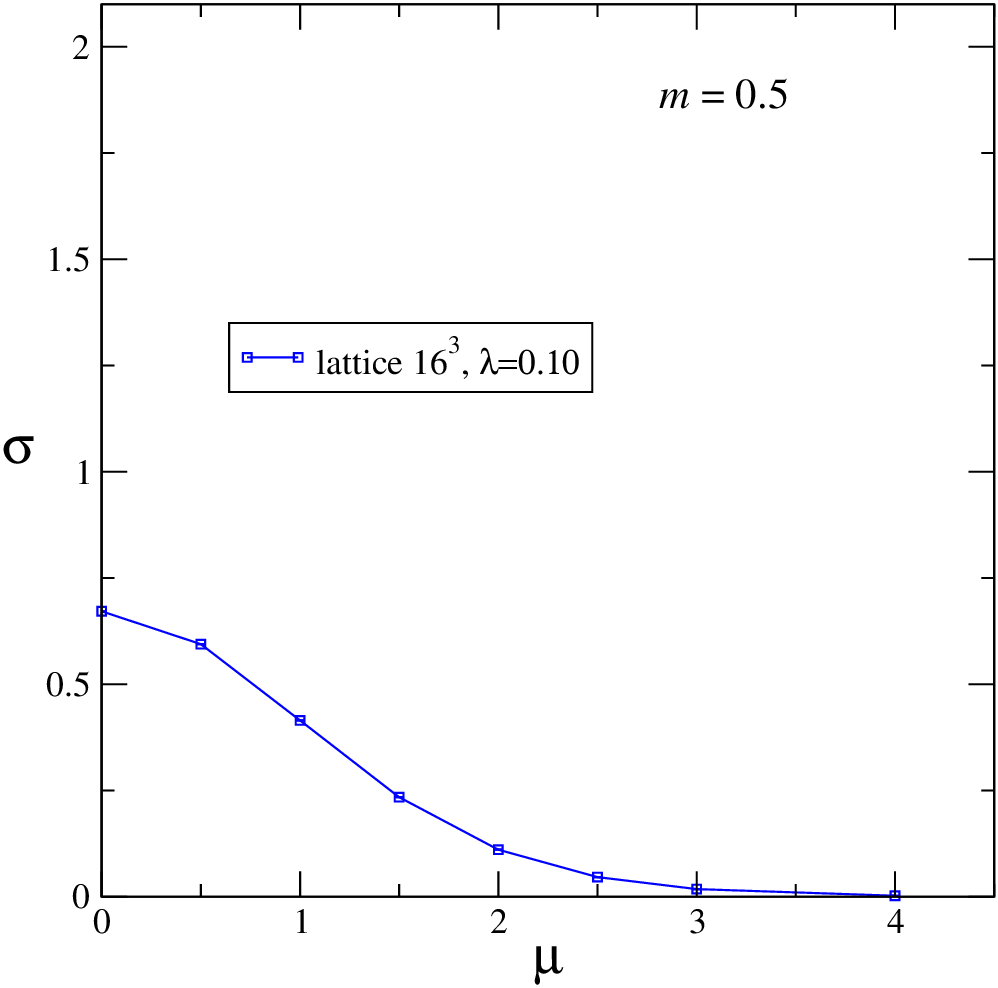}
\hspace{0.3cm}
\includegraphics[width=0.47\linewidth,align=t,clip]{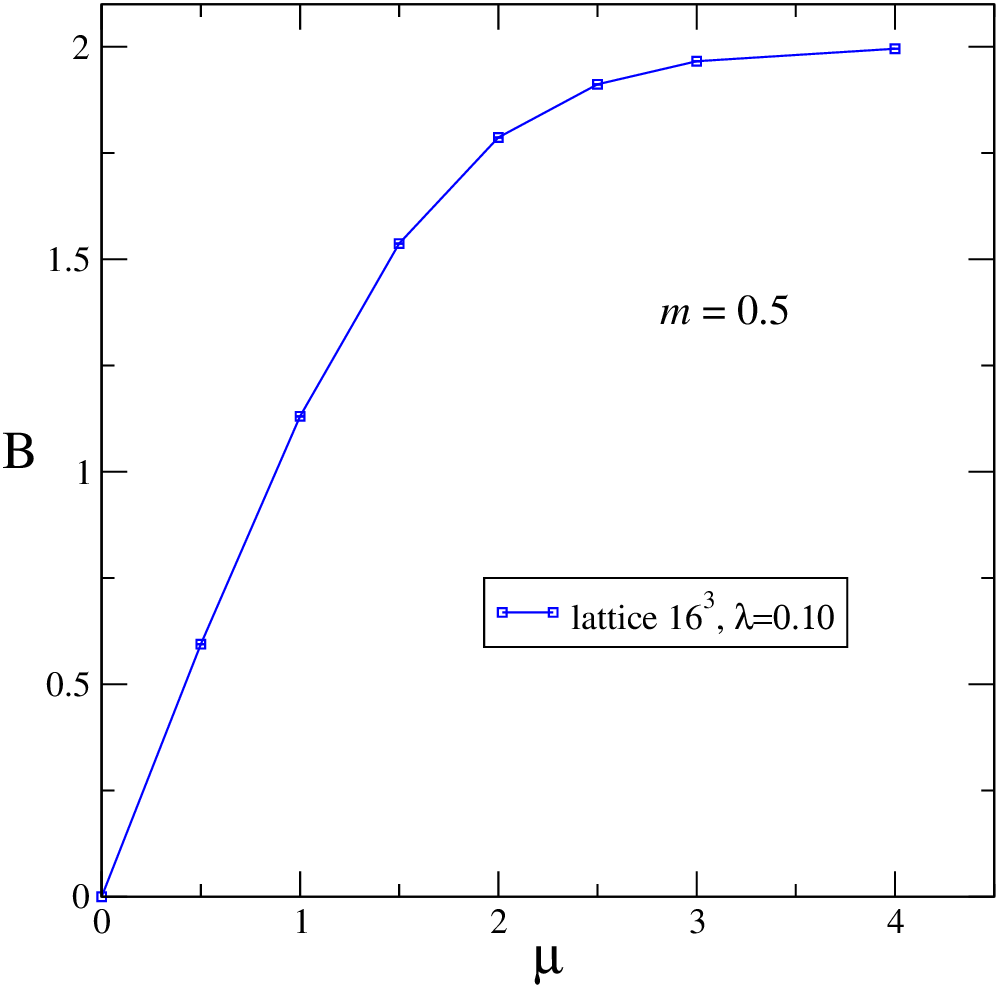}
\caption{(Left) Quark condensate $\sigma$ {\it versus} the chemical potential $\mu$, in the full theory at infinite $N_t$ and
  fermionic mass $m$=0.5, on lattice 16$^3$ for $\lambda$=0.10. (Right) Same as left, for the quark density $B$.
  Solid lines are drawn to guide the eye.}
\label{fig:full_cond_dens_m0.5}
\end{figure}

Numerical results for the condensate $\sigma$ and quark density $B$ show that their dependence on the lattice spatial
volume lies within the numerical uncertainties in the range $L$=8 to $L$=32. We studied their behavior with respect to the
chemical potential $\mu$, in the cases $m=4$ for $\lambda$=0.10 and 0.25 on a lattice $32^3$, and in the case $m=0.5$,
for $\lambda$=0.10 on a lattice $16^3$. The results are shown in Figs.~\ref{fig:full_cond_dens_m4} and~\ref{fig:full_cond_dens_m0.5},
respectively. For $m=4$ the condensate exhibits a smooth transition from a value of about~2 at $\mu$=0 to zero at large $\mu$, with
an inflection point located around $\mu\approx m=4$. The behavior of the quark density is specular, since it starts from zero at $\mu$=0
and saturates at~2 for large $\mu$, showing an inflection point at the same position as for the condensate. Interestingly,
it turns out that the sum of condensate and quark density is almost equal to~2. This feature can be understood
from the analytical point of view in the large-$m$ and/or large-$\mu$ limit. For $m=0.5$ the qualitative behavior of both the condensate
and the quark density are similar, except that the condensate takes a value much smaller than~2 at $\mu=0$. 

\begin{figure}[htb]
\centering
\includegraphics[width=0.47\linewidth,align=t,clip]{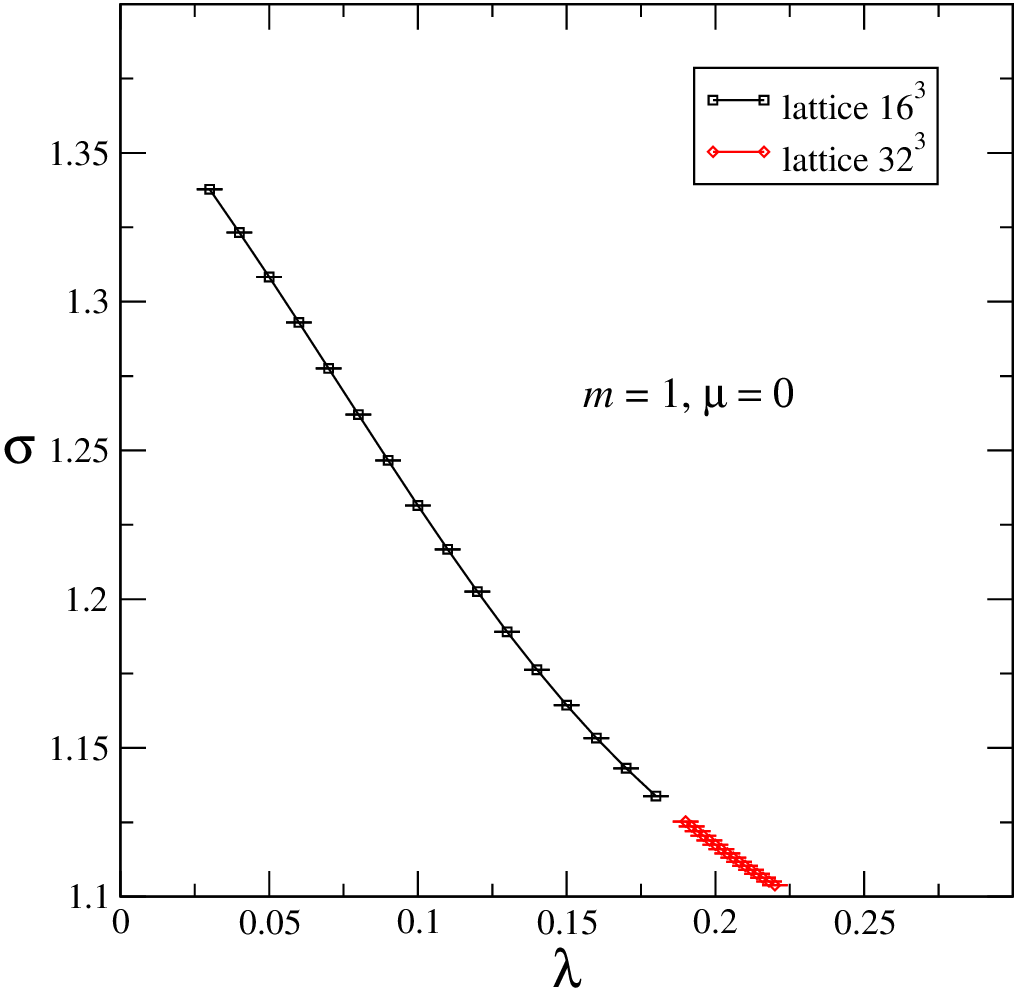}
\hspace{0.3cm}
\includegraphics[width=0.47\linewidth,align=t,clip]{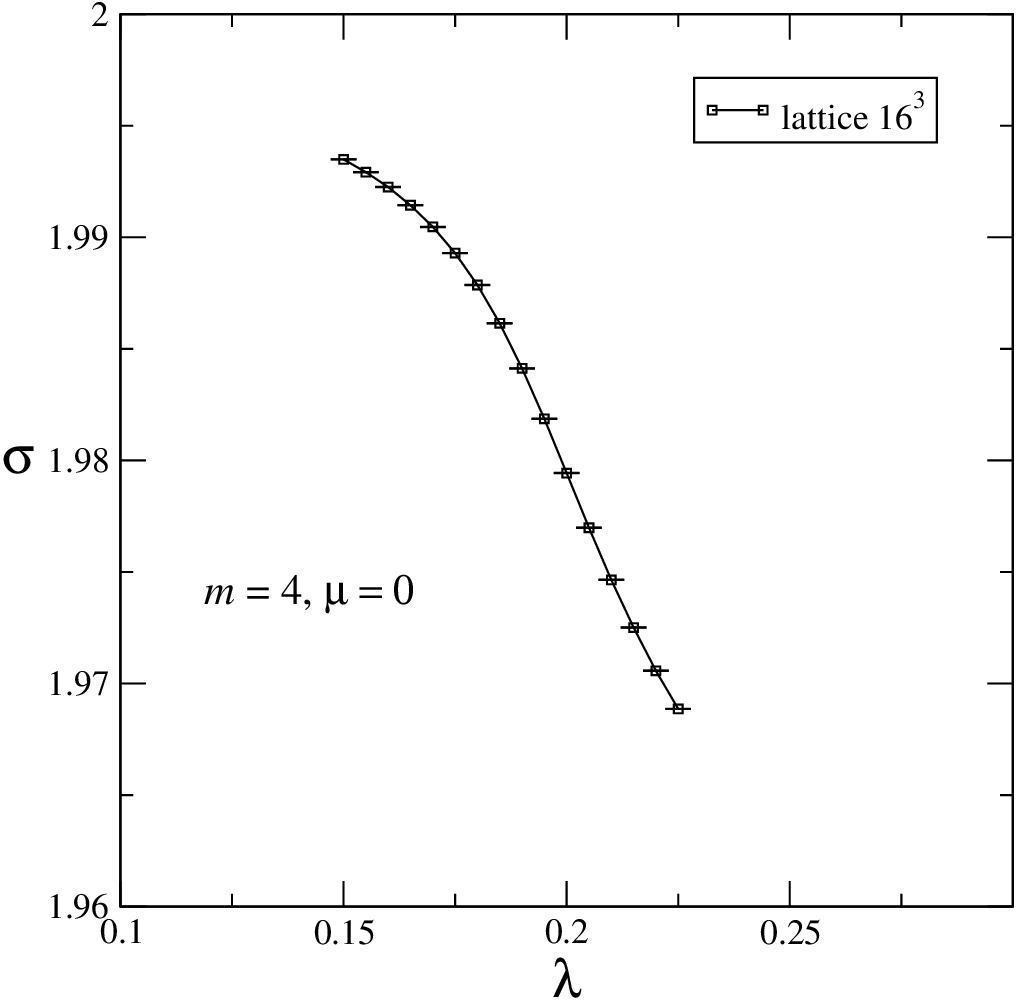}
\caption{Quark condensate $\sigma$ {\it versus} $\lambda$ at zero chemical potential $\mu$ for the fermionic mass $m$=1 (left) and for the fermionic mass $m$=4 (right). Solid lines are drawn to guide the eye.}
\label{fig:quark_cond_m0.0}
\end{figure}

It is also instructive to analyze the behavior of the quark condensate as a function of $\lambda$ at vanishing chemical potential. Fig.~\ref{fig:quark_cond_m0.0} shows such behavior for two values of the fermionic mass $m=1$ and $m=4$ in the vicinity of the crossover. One observes a rapid decrease of the condensate in this region, even though it remains non-zero at all studied $\lambda$ values. This behavior is very similar to the behavior of the quark condensate in $SU(3)$ Polyakov-loop model with the static quark determinant~\cite{Borisenko19}.

\begin{figure}[htb]
\centering
\includegraphics[width=0.47\linewidth,align=t,clip]{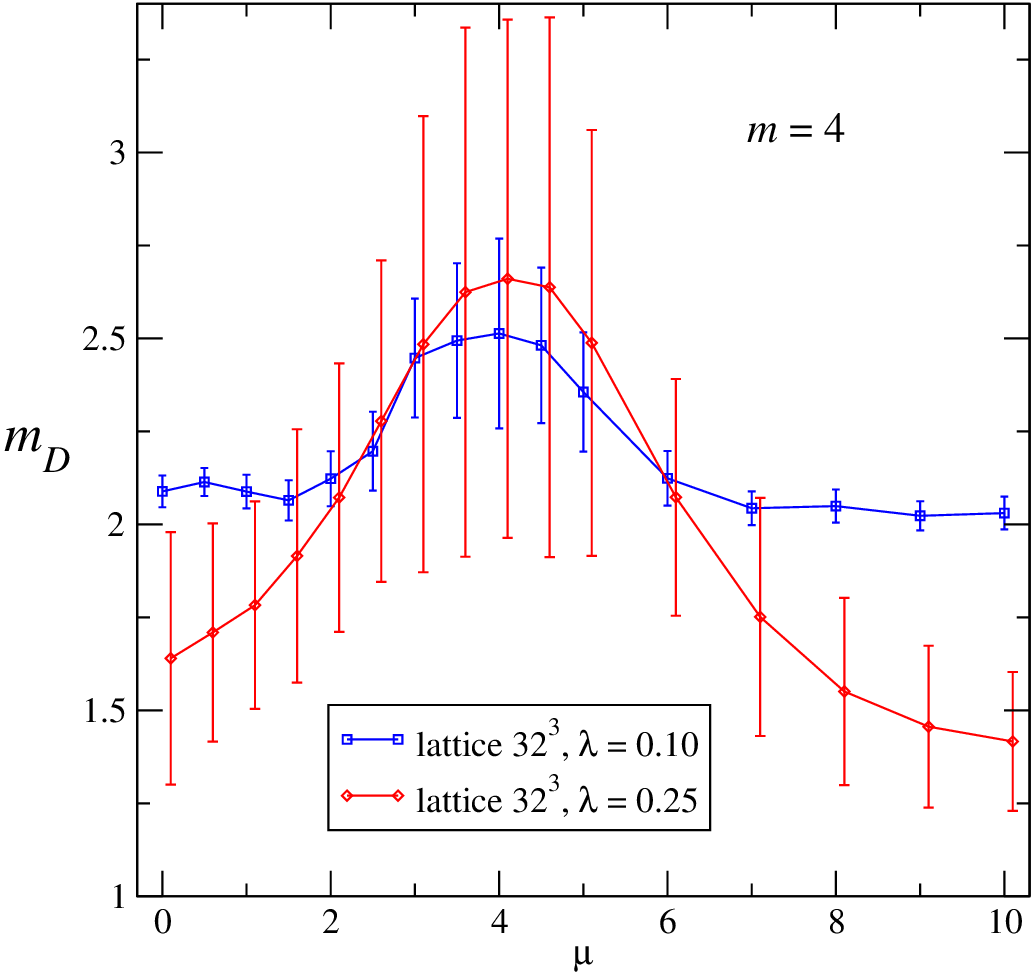}
\caption{Screening mass {\it versus} the chemical potential $\mu$ in the full theory at infinite $N_t$ and
  fermionic mass $m$=4, on lattice 32$^3$ for two values of the coupling $\lambda$.
  The solid lines are drawn to guide the eye.}
\label{fig:full_screening_mass}
\end{figure}

Finally, we considered the behavior of the screening mass $m_D$ {\it versus} $\mu$ for $\lambda$=0.10 and 0.25 and $m$=4 on a lattice $32^3$. We
determined $m_D$ by fitting the correlation data of the Polyakov loop, defined in~(\ref{su2_PL_corr}), with the function
\[
C(r) = A_1 \ \left( \frac{\exp(- m_D r)}{r^{1+c}} + \frac{\exp(-m_D (L-r))}{(L-r)^{1+c}} \right) + A_2  \;,\;\;\;\;\;
r=x_1-y_1\;.
\]
The results of the fits are summarized in Fig.~\ref{fig:full_screening_mass}, which shows a moderate dependence of $m_D$ on $\mu$, except for a small bump around $\mu\approx m=4$. Similar bumps are seen for all sufficiently large quark masses. 
If $m\ll 1$, the maximum of the screening mass is located, presumably at $m=0$ 
(error bars are very large in this case to make a definite conclusion). Such behavior resembles the behavior of the screening mass in 1-dimensional model described in Section~\ref{1d_su2}. 
It is tempting to speculate that such non-monotonic behavior of the screening mass is related to the superfluid phase in $SU(2)$ QCD at large chemical potential. The diquark condensate also exhibits bump-like behavior as $\mu$ grows~\cite{diquark_cond_01}. We cannot check this conjecture within our approximations. We however expect that the screening properties in the superfluid phase are different from the deconfining phase, and this could be seen in the increase of the screening mass. When $\mu>m$, we enter the saturation region and the decrease in the screening mass, and so the disappearance of the bump, could well be the consequence of the saturation.

\clearpage

\section{Summary} 

We showed in this paper that the $SU(N)$ Polyakov-loop model with static quark determinant can be rewritten on a lattice in any dimensions as an Ising model, with the Ising variables attached
to the links. For $N=2$, the Boltzmann weight of that Ising model and all effective couplings between Ising spins have been obtained explicitly. With this formulation, the
model admits a non-zero chemical potential $\mu$ and no sign problem arises. This fact opens the possibility of numerically simulate the model by standard Monte-Carlo algorithms.

Using this formulation we studied the effects of the finite chemical potential in the model. Our main findings can be summarized as follows.

\begin{itemize} 
	\item
	  The 1-dimensional Polyakov-loop model can be solved exactly via the transfer matrix approach. No critical behavior is detected at any values of the model parameters.
          For $\mu$ larger than the quark mass, the quark density exhibits saturation.
	\item 
	  The 3-dimensional pure gauge theory experiences a second order phase transition in the universality class of the 3-dimensional Ising model. This feature agrees with numerous
          previous studies. Among other quantities, we computed the string tension and extracted the critical index $\nu$.  
	\item 
	  The full theory with quarks does not exhibit critical behavior. However, both the quark density and the quark condensate show a rapid change as functions of the chemical
          potential when $\mu$ approaches the quark mass value.
	\item 
	The screening masses $m_D$ show some intriguing behavior as a function of the chemical potential, Fig.~\ref{fig:full_screening_mass}, namely if the quark mass is sufficiently large, $m_D$ shows a bump as a function of the chemical potential centered at $\mu=m$. The nature of this behavior remains an open question.
	
\end{itemize}

\section*{Acknowledgments}

Authors thank V. Chelnokov for many helpful suggestions on different stages of this work. B.A. thanks Juan Jos\'e Alonso for useful discussions and Claudio Bonati for help in the use of the Pisa computer cluster.
B.A., O.B. and A.P. acknowledge support from INFN/Nonperturbative Quantum Chromodynamics (NPQCD) project.
This work is (partially) supported by ICSC – Centro Nazionale di Ricerca in High Performance Computing, Big Data and Quantum Computing, funded by
European Union – NextGenerationEU. S.V. acknowledges support from the Simons Foundation (Grant Number 1039151).

\end{document}